\documentclass[aps,prl,twocolumn,footinbib,superscriptaddress,longbibliography,amsmath,amssymb]{revtex4-2} %preprint

\usepackage{graphicx}
\usepackage{subfigure}
\usepackage{epsfig} 
\usepackage{dcolumn}
\usepackage{bm}
\usepackage{times} %\linenumbers\relax % Commence numbering lines
\usepackage{amsmath}
\usepackage{float}
\usepackage{color}
\usepackage{multirow}
\usepackage[normalem]{ulem}

\begin{document}
\title{Magnetic field-induced ``mirage'' gap in an Ising superconductor}

\author{Gaomin Tang}
%\email{gaomin.tang@unibas.ch}
\affiliation{Department of Physics, University of Basel, Klingelbergstrasse 82, CH-4056
Basel, Switzerland}
\author{Christoph Bruder}
%\email{christoph.bruder@unibas.ch}
\affiliation{Department of Physics, University of Basel, Klingelbergstrasse 82, CH-4056
Basel, Switzerland}
\author{Wolfgang Belzig}
%\email{wolfgang.belzig@uni-konstanz.de}
\affiliation{Fachbereich Physik, Universit\"{a}t Konstanz, D-78457 Konstanz, Germany}

\begin{abstract}
Superconductivity is commonly destroyed by a magnetic field due to orbital or
Zeeman-induced pair breaking. Surprisingly, the spin-valley locking in a two-dimensional
superconductor with spin-orbit interaction makes the superconducting state resilient to
large magnetic fields. 
We investigate the spectral properties of such an Ising superconductor in a magnetic field
taking into account disorder. The interplay of the in-plane magnetic field and the Ising
spin-orbit coupling leads to noncollinear effective fields. We find that the emerging
singlet and triplet pairing correlations manifest themselves in the occurrence of
``mirage'' gaps: at (high) energies of the order of the spin-orbit coupling strength, a
gap-like structure in the spectrum emerges that mirrors the main superconducting gap. We
show that these mirage gaps are signatures of the equal-spin triplet finite-energy pairing
correlations and due to their odd parity are sensitive to intervalley scattering.
\end{abstract}

\maketitle

%\section{Introduction}
{\it Introduction.--}
Superconductivity in two-dimensional materials is a rising topic~\cite{Saito16_review}
since these bear a great potential to host new pairing states due to their high chemical
flexibility and the possibility to combine different materials in van der Waals
heterostructures~\cite{Geim2013}.
Monolayer transition-metal dichalcogenides were recently shown to be two-dimensional
materials~\cite{MoS2_10,MoS2_11,TMD_12,TMD_11} with strong spin-orbit effects. Due to the
broken in-plane inversion symmetry, the spin-orbit coupling arising from the heavy
transition-metal atoms gives rise to a valley-dependent Zeeman-like spin
splitting~\cite{TMD_11,MoS2_12}. Nevertheless, time-reversal symmetry is preserved because
the internal field is opposite in the $K$ and $K'$ valleys. 
Since this Zeeman-like field points out-of-plane, it was termed Ising spin-orbit coupling
(ISOC)~\cite{Zhou16,Lu15,Xi16}. 
In such materials superconductivity has been shown to occur and is believed to be of
$s$-wave type with possible admixtures of triplet pairing channels.

This so-called Ising superconductivity was experimentally realized from the few-layer down
to the monolayer regime in various transition-metal dichalcogenides~\cite{Lu15, Saito16,
Xi16, Xing17, Dvir18,Costanzo18, Lu18, delaBarrera18, Sohn18, Rhodes20, Li20, Cho21,
Kuz21, Hamill21, Kang21, Idzuchi21, Ai21}. 
Since the electrons are confined to a two-dimensional plane, the orbital pair-breaking
effect from an in-plane magnetic field is eliminated~\cite{Tinkham}.
The presence of the ISOC lifts the spin degeneracy in the two valleys and this results in
a considerably enhanced in-plane critical magnetic field~\cite{Bulaevskii76, Gorkov01,
Frigeri04} beyond the Pauli limit~\cite{Chandrasekhar, Clogston}.
Theoretical studies have mainly focused on the phase diagram~\cite{Ilic17, Moeckli18,
Moeckli19, Moeckli20, Moeckli20_magnetic, Moeckli20_DOS, Liu20}, the occurrence of
parity-mixed superconductivity~\cite{Rahimi17, Moeckli18, Moeckli20, IsingSC20_PRX},
% subdominant order parameters~\cite{Moeckli18},
% singlet-triplet mixing~\cite{IsingSC20_PRX},
topological superconductivity~\cite{Zhou16, He2018, Xie20, Ising_CNT}, 
or transport problems~\cite{Zhou16, Transport_Sun18, Transport_Sun19}. 
In particular, the influence of scattering on the $s$-wave gap was
investigated~\cite{Ilic17, Moeckli20, Moeckli20_magnetic}.
Moreover, due to the ISOC an in-plane magnetic field can mediate the conversion from
singlet Cooper pairs to equal-spin triplet pairs~\cite{Rahimi17, Moeckli18, Moeckli19,
Moeckli20, Moeckli20_DOS}.

\begin{figure*}
\centering
\includegraphics[width=5.5in]{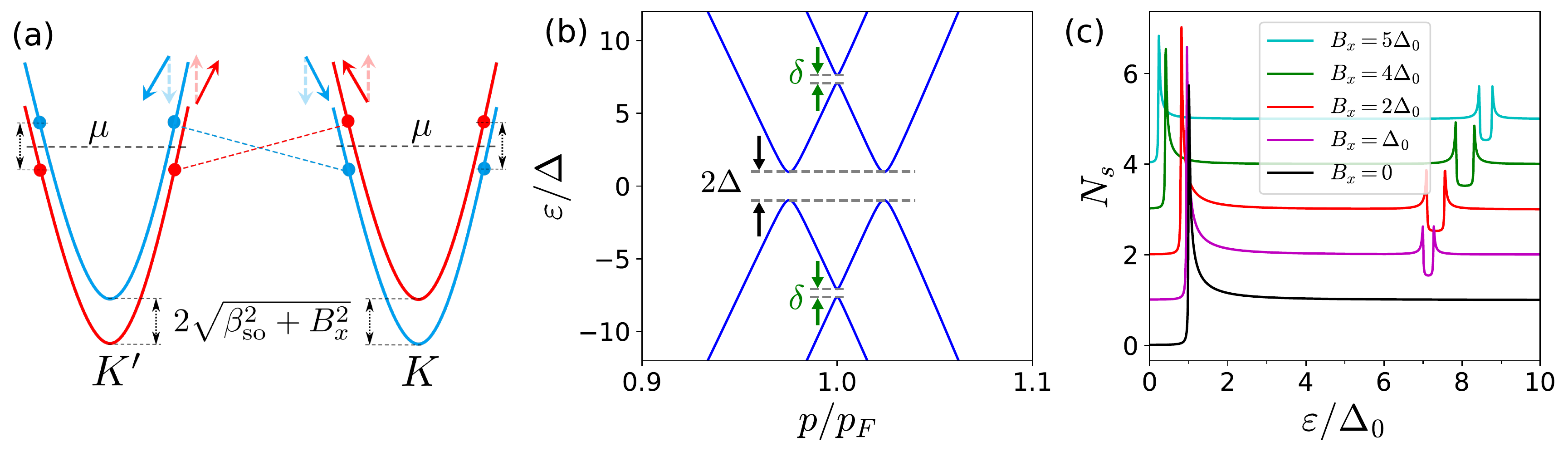}
\caption{(a) Schematic band structure in the normal state. The electrons near the $K$ and
  $K'$ valleys are subject to the ISOC $\beta_{\rm so}$, which pins the electron
  spins to the out-of-plane direction (dashed arrows), and an in-plane magnetic field
  $B_x$.
  For finite $B_x$, the spin directions are reoriented (solid arrows)
  %\sout{and the band splitting in both valleys is $2\sqrt{\beta_{\rm so}^2+B_x^2}$}.
  (b) Quasiparticle energy spectrum of Eq.~\eqref{H_BdG} near the Fermi momentum $p_F$
  with $\beta_{\rm so}=7\Delta$, $B_x=2\Delta$ and $\mu=150\Delta$. 
  %\sout{A similar gap structure occurs for negative $p$ (not shown).} 
  The \textit{mirage} gaps $\delta$ are shifted images of the main superconducting gap. 
  (c) Density of states $N_s$ for different $B_x$ in the clean limit. All lines for
  finite $B_x$ have been offset for better visibility. 
  %\sout{Since $N_s$ is even in energy, the negative energy parts are not shown.} 
  Here, $\beta_{\rm so}=7\Delta_0$ and $T=0.1T_{c0}$, where $\Delta_0$ and $T_{c0}$ are,
  respectively, the zero-temperature gap and transition temperature in the absence of a
  magnetic field. }
\label{fig1}
\end{figure*}

In this Letter, we discuss the emergence of finite-energy pairing correlations in an Ising
superconductor subject to an in-plane magnetic field. 
%in a quantitative way using the quasiclassical Green's function formalism. 
We show that these correlations are accompanied
by the appearance of two symmetric mirages of the main superconducting gap shifted to a
finite energy [See Figs.~\ref{fig1}(b) and \ref{fig1}(c)].
%that is determined by the ISOC 
This picture is confirmed by relating the mirage gaps to finite-energy pairing that
results from a subtle interplay between noncollinear spins and the valley degree of
freedom. Using a fully self-consistent approach, we show that the intervalley scattering
due to nonmagnetic impurities destroys the mirage gaps. 
%while the standard $s$-wave mixed-spin superconductivity survives due to Anderson's
%theorem.

{\it Hamiltonian.--}
For an Ising superconductor with a spin-singlet $s$-wave pairing gap $\Delta$, the
effective Bogoliubov-de Gennes Hamiltonian near one of the valleys can be written in the
Nambu basis $(c_{{\bm k},\uparrow}, c_{{\bm k},\downarrow}, c_{-{\bm k},\uparrow}^\dag,
c_{-{\bm k},\downarrow}^\dag)$ as
\begin{equation} \label{H_BdG}
  H_{\mathrm{BdG}} = 
  \begin{bmatrix}
    H_0({\bm k})  &  \Delta i\sigma_y  \\
    -\Delta i\sigma_y  &  -H_0^*(-{\bm k}) 
  \end{bmatrix} .
\end{equation}
Here, $H_0$ is 
\begin{equation}
  H_0({\bm k}={\bm p}+s {\bm K})=\xi_{\bm p}\sigma_0 +s\beta_{\rm so} \sigma_z - B_x
  \sigma_x ,
\end{equation}
where $s{\bm K}$ is the position of the valley $K$ ($s=+$) or $K'$ ($s=-$) in momentum
space and ${\bm p}$ is the deviation from the ${\bm K}({\bm K'})$-point. Furthermore,
$\xi_{\bm p}=p^2/(2m)-\mu$ is the dispersion measured form the chemical potential $\mu$. 
The Pauli matrices $\sigma_x$, $\sigma_y$, and $\sigma_z$ act on the spin space and
$\sigma_0$ is the unit matrix. The ISOC $\beta_{\rm so}$ pins the electron spins
to the out-of-plane. 
The in-plane magnetic field $B_x$ is along the $x$-direction and induces the Zeeman term
$-B_x\sigma_x$. The prefactor $g_L\mu_B/2$ with the Land\'{e}
g-factor $g_L$ and the Bohr magneton $\mu_B$ has been absorbed in $B_x$. 
Since the magnetic field $B_x$, which is valley symmetric, tends to tilt the electron
spins in the $x$-direction, the spins are reoriented [See Fig.~\ref{fig1}(a)]. The band
splitting in the normal state is $2\sqrt{\beta_{\rm so}^2+B_x^2}$. 

{\it Finite-energy pairing.--}
The general pairing-correlation function can be expressed as~\cite{Gorkov01, Frigeri04}
\begin{equation} \label{correlation}
  F({\bm k}, \varepsilon) = \Delta\big[ F_0({\bm k}, \varepsilon)\sigma_0 
    + {\bm F}({\bm k}, \varepsilon)\cdot{\bm \sigma} \big] i\sigma_y ,
\end{equation}
where $F_0$ and ${\bm F}$, respectively, parametrize the singlet and triplet pairing
correlations~\cite{SM}. 
%Explicit expressions for $F_0$ and ${\bm F}$ can be obtained from the
%Gor'kov equation and are given in the Supplemental Material~\cite{SM}. 
Using Eq.~\eqref{correlation}, the pairing wave function can be written as
\begin{align}
  |\Psi\rangle 
  = &F_0 \big( |{\uparrow}{\downarrow}\rangle -|{\downarrow}{\uparrow}\rangle \big)
  + F_x \big( |{\downarrow}{\downarrow}\rangle -|{\uparrow}{\uparrow}\rangle \big) \notag \\
  +i &F_y \big( |{\downarrow}{\downarrow}\rangle +|{\uparrow}{\uparrow}\rangle \big) 
  + F_z \big( |{\uparrow}{\downarrow}\rangle +|{\downarrow}{\uparrow}\rangle \big) 
  \label{Psi}.
\end{align}
Here, the momentum dependence is omitted, for example, $|{\bm k}{\uparrow}, -{\bm
k}{\downarrow} \rangle$ is abbreviated as $|{\uparrow}{\downarrow} \rangle$. 

We first discuss the low-energy pairing that occurs around the Fermi energy. In the
absence of a magnetic field $B_x$, the ISOC field results in opposite energy
splittings in the two valleys so that the amplitude of the pairing state
$|{\uparrow}{\downarrow}\rangle$ is different from that of
$|{\downarrow}{\uparrow}\rangle$ except at the Fermi momentum. Hence, in addition to the
standard singlet pair amplitude $\propto F_0$, a pairing state
$|{\uparrow}{\downarrow}\rangle +|{\downarrow}{\uparrow}\rangle$ is created,
i.e., $F_z$ is finite~\cite{Gorkov01,Frigeri04,Zhou16,Moeckli20}. This pair
amplitude $F_z$, which is due to the ISOC, has the form $F_z \propto s\beta_{\rm so}
\xi_{\bm p}$~\cite{Zhou16,Rahimi17,SM} and is odd in the valley index. 
% $F_z$ is odd in valley, since it results from the in-plane inversion symmetry breaking.
In the presence of $B_x$, the electron spins are reoriented so
%an in-plane magnetic field 
that equal-spin pairing states $|{\uparrow}{\uparrow}\rangle_x$ and
$|{\downarrow}{\downarrow}\rangle_x$ emerge around the Fermi energy.
Here, the subscript `$x$' denotes the spin states in the $x$-direction.
This leads to the triplet states
$|{\uparrow}{\uparrow}\rangle_x + |{\downarrow}{\downarrow}\rangle_x$ and 
$|{\uparrow}{\uparrow}\rangle_x - |{\downarrow}{\downarrow}\rangle_x$,
which in the $z$-basis take the form
$|{\uparrow}{\uparrow}\rangle + |{\downarrow}{\downarrow}\rangle$ and
$|{\uparrow}{\downarrow}\rangle + |{\downarrow}{\uparrow}\rangle$,
respectively~\cite{SM}.

The interplay between ISOC and an in-plane magnetic field leads to a new feature,
viz., finite-energy pairing correlations. A qualitative illustration is provided in
Fig.~\ref{fig1}(a) that shows the schematic band structure with electrons at Fermi momenta
$p_F=\pm\sqrt{2m\mu}$ as blue and red dots.
%\sout{In each valley, these blue and red dots are separated by an energy of
%$2\sqrt{\beta_{\rm so}^2+B_x^2}$ in the normal state.}
Near the Fermi momentum, the electron at
$|{\bm k}{\uparrow}\rangle_x$ ($|{\bm k}{\downarrow}\rangle_x$)
can pair with the electron at $|-{\bm k}{\downarrow}\rangle_x$
($|-{\bm k}{\uparrow}\rangle_x$) as indicated by the dashed lines in Fig.~\ref{fig1}(a).  
As a consequence, there is a coexistence of the singlet state
$|{\uparrow}{\downarrow}\rangle_x - |{\downarrow}{\uparrow}\rangle_x$ and the triplet
state $|{\uparrow}{\downarrow}\rangle_x + |{\downarrow}{\uparrow}\rangle_x$, which in the
$z$-basis is $|{\downarrow}{\downarrow}\rangle - |{\uparrow}{\uparrow}\rangle$~\cite{SM}. 
Similarly, there are also equal-spin triplet states $|{\downarrow}{\downarrow}\rangle$ and
$|{\uparrow}{\uparrow}\rangle$ near the Fermi momenta in the $z$-direction. 
This can give rise to the equal-spin triplet states $|{\downarrow}{\downarrow}\rangle
-|{\uparrow}{\uparrow}\rangle$ and $|{\downarrow}{\downarrow}\rangle
+|{\uparrow}{\uparrow}\rangle$. 
We term these finite-energy pairing states, since the two electrons forming a Cooper pair
have opposite energies with respect to the Fermi energy and are separated in energy by
about $2\sqrt{\beta_{\rm so}^2+B_x^2}$, which is typically much larger than $2\Delta$.

The pairing states and the symmetries of the corresponding pair amplitudes are summarized
in Table~\ref{tab:1}. The pair amplitude $F_x$ is odd in time, since $B_x$ breaks the
time-reversal symmetry. The overall antisymmetry of the Cooper pair wave function is
ensured by the parity under exchanging the arguments of spin, valley, and
time~\cite{Berezinskii74,Tanaka12,Tanaka16, RMP19_oddSC}. The symmetries of the amplitudes
$F_x$ and $iF_y$ that are even and odd, respectively, under exchanging the valley indices,
will become clear later in Eq.~\eqref{fxy} from the quasiclassical Green's function
formalism.

\begin{table}[t] 
  \begin{center} 
    \begin{tabular}{|c||c|c|c|c|c|}
      \hline
       & pairing states & pairing states & \multirow{2}{*}{spin} & \multirow{2}{*}{valley}
       & \multirow{2}{*}{time} \\
       & (zero energy) & (finite energy) & & & \\
      \hline 
      $F_0$ & $|{\uparrow}{\downarrow}\rangle - |{\downarrow}{\uparrow}\rangle$ 
      & $|{\uparrow}{\downarrow}\rangle - |{\downarrow}{\uparrow}\rangle$
      & singlet & even & even  \\
      \hline
      $F_x$ & $\times$
      & $|{\downarrow}{\downarrow}\rangle - |{\uparrow}{\uparrow}\rangle$ 
      & triplet & even & odd  \\
      \hline
      $iF_y$ & $|{\downarrow}{\downarrow}\rangle + |{\uparrow}{\uparrow}\rangle$ 
      & $|{\downarrow}{\downarrow}\rangle + |{\uparrow}{\uparrow}\rangle$ 
      & triplet & odd & even \\
      \hline
      $F_z$ & $|{\uparrow}{\downarrow}\rangle + |{\downarrow}{\uparrow}\rangle$ & 
      $\times$ & triplet & odd & even \\
      \hline
    \end{tabular}
    \caption{\label{tab:1}Pairing states at zero and finite energy of an Ising
      superconductor subject to an in-plane magnetic field.
      Zero-energy $F_x$ pairing and finite-energy $F_z$ pairing do not exist. The
      symmetries of the pair amplitudes are characterized by the parity under exchanging
      the arguments of spin, valley, and time. }
  \end{center}
\end{table}

The presence of the finite-energy pairing correlations is reflected in the density of
states (DOS). The quasiparticle energy spectrum for $\Delta \ll \beta_{\rm so} \ll \mu$ is
shown in Fig.~\ref{fig1}(b). In addition to the main superconducting gap, there
are mirage gaps of size $\delta$ appearing at the Fermi momentum.
These mirage gaps can be interpreted as an image of the main superconducting gap shifted
by the effective field and is a hallmark of the finite-energy pairing correlations.
Note that the DOS is finite in the mirage gaps, since in each gap only one band in each
valley participates in the finite-energy pairing [See Fig.~\ref{fig1}(a)]. The mirage gaps
are located at $\pm\varepsilon_0$ with $\varepsilon_0 = (\varepsilon_1 + \varepsilon_2
)/2$, where $\varepsilon_{1(2)}=\sqrt{\beta_{\rm so}^2+(B_x\pm \Delta)^2}$ are the
eigenvalues of the Hamiltonian $H_{\rm BdG}$ at $\xi_{\bm p}=0$. Their widths are $\delta
= \varepsilon_1-\varepsilon_2$. 
The location and width of mirage gap can be used to experimentally extract the
strength of the ISOC. This is necessary to estimate the upper critical magnetic fields
at low temperatures, which are too large to be measured directly at present.
For the case without ISOC ($\beta_{\rm so}=0$), we arrive at $\varepsilon_0
=\pm\Delta$ and $\delta = 2B_x$ and this reduces to the well-known Zeeman splitting
between the spin-up and spin-down electrons. This splitting suppresses the formation of
Cooper pairs and results in the paramagnetic limit of
superconductivity~\cite{Chandrasekhar, Clogston}.
For $\beta_{\rm so}\gg \Delta$, the mirage gaps are clearly separated from the main gaps;
they appear around the energy
\begin{equation} \label{ve0}
  \varepsilon_0 \approx \sqrt{\beta_{\rm so}^2 + B_x^2} ,
\end{equation}
and have widths
\begin{equation} \label{delta}
  \delta \approx 2 \Delta B_x/\sqrt{\beta_{\rm so}^2 + B_x^2}.
\end{equation}
It can be inferred that in the absence of $B_x$, $\delta$ vanishes, i.e., there is no
finite-energy pairing. Note that for $ B_x\gtrsim \beta_{\rm so}$ pair breaking
sets in so that the main superconducting gap vanishes and consequently so do the
mirage gaps.

Figure~\ref{fig1}(c) shows the superconducting DOS $N_s$ (normalized to that of the normal
state) for different in-plane magnetic fields in the clean limit. The curves are
calculated using the quasiclassical Green's function formalism.
The DOS is $0.5$ inside each mirage gap, since only one band in each valley participates
in the pairing. On increasing the magnetic field, the superconducting gap $\Delta$
decreases, while the mirage gap $\delta$ shows a nonmonotonic behavior and first increases
and then decreases. 
This is due to the interplay between the gap $\Delta$ and the magnetic field $B_x$, as
indicated by Eq.~\eqref{delta}. Below a certain value of $B_x$, the decreasing slope of
$\Delta$ is smaller than the increasing slope of $B_x$ [See Figs.~S4(b) and S4(d) in the
Supplemental Material~\cite{SM}], so that the increase of $B_x$ dominates. However, above
this value of $B_x$, the decrease of $\Delta$ dominates and $\delta$ decreases.

%\section{Quasiclassical Green's function}
{\it Quasiclassical Green's function.--}
We now describe the formalism used to calculate the
DOS and the pair amplitudes. Since the quasiclassical formalism concentrates on
the phenomena close to the Fermi surface~\cite{Eilenberger1968, LO1969, Belzig99,
noneqSC}, it can be applied to the situation where both the superconducting gap and the
ISOC are much smaller than the Fermi energy. 
%\sout{This condition is satisfied in ion-gated MoS$_2$ where the critical temperature is
%about $6.5\,$K, the ISOC strength $\beta_{\rm so}\approx 13\,$meV and Fermi energy
%$\mu\approx 150\,$meV~\cite{Saito16_review}.  (Ref.~\cite{Lu18} reported $\beta_{\rm
%so}=6.2\,$meV in MoS$_2$.) }
The general structure of the quasiclassical Green's function in Nambu space is
\cite{noneqSC, Eschrig15}
\begin{equation}
  \hat{g}(\hat{\bm k}, \varepsilon) =
  \begin{bmatrix}
    g_0\sigma_0 + \bm{g}\cdot \bm{\sigma} & (f_0\sigma_0 + \bm{f}\cdot \bm{\sigma}) 
    i\sigma_y \\
    (\bar{f}_0\sigma_0 + \bar{\bm{f}} \cdot \bm{\sigma}^*)i\sigma_y  & \bar{g}_0
    \sigma_0+\bar{\bm{g}}\cdot {\bm \sigma}^* 
  \end{bmatrix} ,
\end{equation}
where $\hat{\bm k}$ denotes the direction of momentum ${\bm k}$ and $\varepsilon$ is
the quasiparticle energy with respect to the Fermi energy.
The bar operation is defined as $\bar{q}(\hat{\bm k}, \varepsilon) = q(-\hat{{\bm k}},
-\varepsilon^*)^*$ with $q\in \{ g_0, f_0, {\bm g}, {\bm f} \}$. 
The anomalous Green's functions $f_0$ and $\bm{f}$, respectively, correspond to $F_0$ and
${\bm F}$ in Eq.~\eqref{correlation}. The DOS $N_s$ is given by ${\rm Re}(g_0)$.
%, i.e., $N_s\equiv {\rm Re}(g_0)$. 

The Eilenberger equation for a homogeneous system reads~\cite{Eilenberger1968, noneqSC}
\begin{equation}
  \big[\varepsilon \sigma_0\tau_3 -\hat{\Delta}-\hat{\nu} -\hat{\Sigma}(\varepsilon),\,
  \hat{g} \big] =0, 
\end{equation}
with the order parameter term $\hat{\Delta} =\Delta i\sigma_y \tau_2$. 
The Pauli matrices $\tau_1$ $\tau_2$, and $\tau_3$ act on the Nambu space and $\tau_0$ is
the corresponding unit matrix. The ISOC and Zeeman terms are included in $\hat{\nu}$ with
$\hat{\nu} = s\beta_{\rm so}\sigma_z \tau_3 -B_x \sigma_x \tau_0$.
Nonmagnetic impurities are taken into account using the self-consistent Born approximation
with
$\hat{\Sigma}(\varepsilon) = -i\Gamma\langle \hat{g}(\hat{\bm k}, \varepsilon)\rangle$,
where $\Gamma$ is the intervalley impurity scattering rate and $\langle \cdots \rangle$
denotes the average over the Fermi momentum direction. 
It has been theoretically demonstrated that nonmagnetic intervalley scattering can
suppress the upturn of the in-plane critical magnetic field in the low temperature
region~\cite{Ilic17, Moeckli20}.
According to Anderson's theorem~\cite{Anderson_theorem}, intravalley nonmagnetic
scattering has no effect for an $s$-wave superconductor, which is the case here. 
By combining with ${\rm Tr}(\hat{g}) =0$ and the normalization condition $\hat{g} \hat{g}
=\sigma_0\tau_0$, all the components of $\hat{g}$ can be obtained. 
%Since only $s$-wave pairing is considered, the components of $\hat{g}(\hat{\bm k},
%\varepsilon)$ are independent of momentum direction $\hat{\bm k}$ in a given valley. 
In particular, $f_x$ and $f_y$ can be, respectively, written as
\begin{equation} \label{fxy}
  f_x = a\ \tilde{\varepsilon} B_x , \qquad f_y = a\ is\beta_{\rm so} B_x ,
\end{equation}
where $\tilde{\varepsilon}=\varepsilon +i\Gamma g_0$ and $a$ is fixed by the normalization
condition. 
The derivation and calculation details can be found in the Supplemental
Material~\cite{SM}. 
%\sout{The different components have the following interpretation:
%${\rm Re}(g_{+,x})$ gives the total magnetization of both valleys in the $x$-direction and
%${\rm Re}(g_{-,z})$ the difference of the spin polarizations between the $K$ and $K'$
%valleys in the $z$-direction. The term $f_0$ characterizes the singlet pairing and the
%terms $f_x$ and $if_y$ the triplet pairing correlations.}
Equation~\eqref{fxy} shows that the $if_y$ pairing is a consequence of the interplay
between ISOC and the in-plane magnetic field. We can also deduce from Eq.~\eqref{fxy}
that $f_x$ and $f_y$ are even and odd with respect to the valley index $s$, respectively.
This confirms the valley symmetries shown in Table~\ref{tab:1}. 
The amplitudes of $f_x$ and $if_y$ are not equal around $\varepsilon =\pm\varepsilon_0$
and the difference lies in the finite-energy pairing $|{\uparrow}{\downarrow}\rangle_x
+|{\downarrow}{\uparrow}\rangle_x$. 
Since only an in-plane magnetic field is applied and the dispersion $\xi_{\bm p}$ is
integrated over in the quasiclassical formalism~\cite{noneqSC,Moeckli18}, the
quasiclassical pair amplitude $f_z$ vanishes.

\begin{figure}
\centering
\includegraphics[width=\columnwidth]{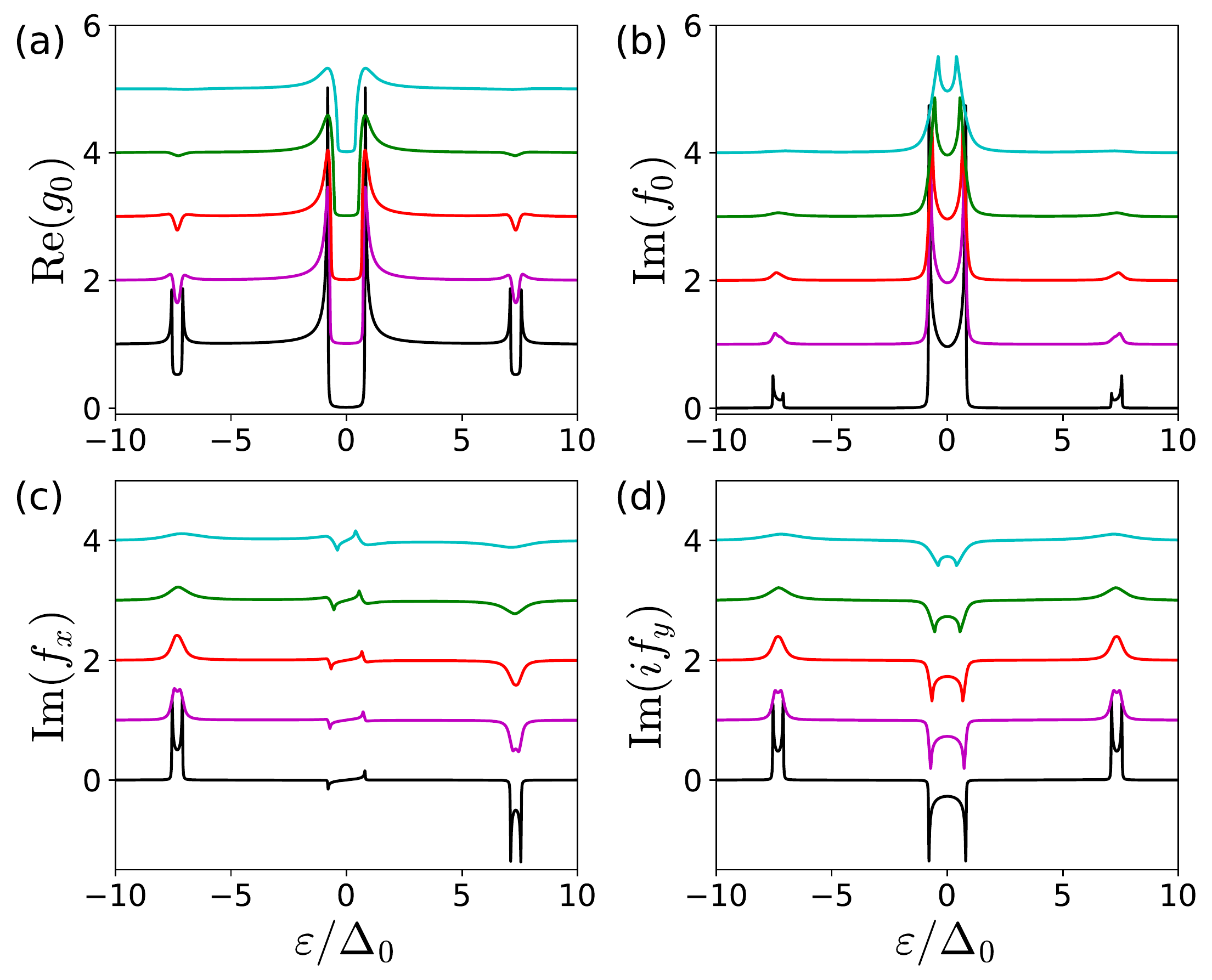}
\caption{Effects of intervalley scattering. Various components of the Green's function
  $\hat{g}$ at different intervalley scattering strengths (from bottom to top:
  $\Gamma=0$, $0.2\Delta_0$, $0.4\Delta_0$, $\Delta_0$, $2\Delta_0$) in valley $K$ with
  $\beta_{\rm so}=7\Delta_0$, $B_x=2\Delta_0$ and $T=0.1T_{c0}$.
  %where $\Delta_0$ and $T_{c0}$ are, respectively, the zero-temperature gap and the
  %transition temperature in the absence of a magnetic field. 
  All curves for finite $\Gamma$ have been offset for better visibility. 
  %The presented results are for the $K$ valley; 
  For valley $K'$, the sign of $f_y$ reverses, while the signs of $g_0$, $f_0$ and $f_x$
  remain unchanged.}
\label{fig2}
\end{figure}

%\section{Numerical Results and discussion}
{\it Numerical results and discussion.--}
To unveil the microscopic mechanism, we present self-consistent numerical results at
different intervalley scattering strengths $\Gamma$ in Fig.~\ref{fig2}. 
To account for inelastic processes, a Dynes broadening parameter $\eta=0.01\Delta_0$ has
been added, ${\varepsilon}{\rightarrow\,}{\varepsilon} +i\eta$~\cite{Dynes}. 
Figure~\ref{fig2}(b) shows that $\textrm{Im}(f_0)$ is finite near $\varepsilon
=\pm\varepsilon_0$; this is the consequence of finite-energy singlet pairing
$|{\uparrow}{\downarrow}\rangle_x-|{\downarrow}{\uparrow}\rangle_x$. Finite-energy
equal-spin pairing correlations $|{\downarrow}{\downarrow}\rangle \mp
|{\uparrow}{\uparrow}\rangle$ are visible in Figs.~\ref{fig2}(c) and \ref{fig2}(d).
%Regarding the momentum ${\bm k}$ residing in the $K$ valley, 
An electron in $K$ valley in state $|{\bm k}{\uparrow}\rangle$ near energy
$\varepsilon=\varepsilon_0$ pairs with the electron in state
$|-{\bm k}{\uparrow}\rangle$ with energy $-\varepsilon$ forming the pairing state
$|{\uparrow}{\uparrow}\rangle$ [See Fig.~\ref{fig1}(a)].
Consequently, ${\rm Im}(f_x)$ is negative while ${\rm Im}(if_y)$
is positive around $\varepsilon =\varepsilon_0$. Similarly, ${\rm
Im}(f_x)$ and ${\rm Im}(if_y)$ are positive around $\varepsilon =-\varepsilon_0$ due to
the pairing $|{\downarrow}{\downarrow}\rangle$. 
The pair amplitude ${\rm Im}(f_x)$ is odd in energy due to the time-reversal symmetry
breaking induced by the in-plane magnetic field. 
The finite values of ${\rm Im}(if_y)$ around the Fermi energy ($\varepsilon=0$) are a
manifestation of the zero-energy pairing $|{\uparrow}{\uparrow}\rangle_x
+|{\downarrow}{\downarrow}\rangle_x$. The finite values of $f_x$ near $\varepsilon=\pm
\Delta$ are due to the Dynes broadening used in the numerical calculation.

%To unveil the microscopic mechanism of the finite-energy pairing, we present
%self-consistent numerical results using parameters $\beta_{\rm so}
%=7\Delta_0$, $B_x =2\Delta_0$, and $T=0.1T_{c0}$, where $\Delta_0$ and $T_{c0}$ are,
%respectively, the zero-temperature gap and the transition temperature in the absence of a
%magnetic field. 
%To account for inelastic processes, a Dynes broadening parameter
%$\eta=0.01\Delta_0$ has been added, ${\varepsilon}{\rightarrow\,}{\varepsilon}
%+i\eta$~\cite{Dynes}. Various components of the Green's function at different intervalley
%scattering strengths $\Gamma$ are shown in Fig.~\ref{fig2}. 
%where the clean limit is shown as black lines. 

%We first focus on the clean limit. For the DOS $N_s\equiv {\rm Re}(g_0)$, the black line
%in Fig.~\ref{fig2}(a) corresponds to the red line in Fig.~\ref{fig1}(c). 
%Because of the Zeeman splitting caused by the ISOC and the magnetic field, both ${\rm
%Re}(g_{+,x})$ and ${\rm Re}(g_{-,z})$ are odd in energy with 
%\begin{equation}
%  {\rm Re}(g_{+,x})\approx \pm \frac{B_x^2}{\beta_{\rm so}^2+B_x^2} , \ \
%  {\rm Re}(g_{-,z})\approx \mp \frac{\beta_{\rm so}^2}{2(\beta_{\rm so}^2+B_x^2)},
%\end{equation}
%around $\varepsilon =\pm\varepsilon_0$. 

We now turn to the discussion of nonmagnetic intervalley scattering effects. 
As can be seen from the DOS [Fig.~\ref{fig2}(a)] and the singlet pair amplitude
[Fig.~\ref{fig2}(b)], the superconducting gap decreases with increasing impurity
scattering strength~\cite{Ilic17, Moeckli20}. Meanwhile, the finite-energy
pairing correlations get suppressed as well and are more sensitive to the impurity
scattering than the zero-energy singlet pairing. 
Because of the suppression of the finite-energy pairing correlations, the DOS inside the
mirage gaps increases. 
It can be seen from Figs.~\ref{fig2}(c) and \ref{fig2}(d) that the finite-energy
pairing correlations almost vanish for $\Gamma=2\Delta_0$. 
The effect of nonmagnetic intervalley scattering can be explained as follows.
First, the mirage gap is proportional to the main superconducting gap
which is suppressed due to the intervalley scattering. 
Since nonmagnetic scattering is spin-conserving, it connects electron states from
different valleys with the same out-of-plane spin direction. Due to the spin
reorientation by the in-plane magnetic field, there is scattering between 
$|{\bm k}{\uparrow}\rangle_x$ and $|{-}{\bm k}{\downarrow}\rangle_x$ which in turn 
reduces the effect of the magnetic field.
The effective magnetic field becomes $\tilde{B}_x=B_x+i\Gamma g_{+,x}$ with $g_{+,x}$
characterizing the in-plane magnetization induced by $B_x$~\cite{SM}. 
This further reduces the mirage gap width. 
Moreover, because of the finite DOS in one valley inside the mirage gap of the other
valley, nonmagnetic impurity scattering leads to an imaginary part of the energy
$\sim\Gamma g_0 \gtrsim \Gamma/2$, so that the coherence peaks of the mirage gaps are
smeared. A more detailed analytical treatment of the impurity effect is presented in the
Supplemental Material~\cite{SM}.
%The suppression of the finite-energy pairing correlations occurs since the Zeeman effect
%induced by $B_x$ is reduced by the impurity scattering between $|{\bm
%k}{\uparrow}\rangle_x$ and $|-{\bm k}{\downarrow}\rangle_x$. 
%%This is reflected in the amplitude of $\mathrm{Re}(g_{+,x})$
%%that decreases with increasing $\Gamma$. 
%Once $\Gamma$ is comparable to $B_x$, the finite-energy pairing correlations almost
%vanish. 
%According to Ref.~\cite{Ilic17}, $\Gamma\approx 0.4\Delta_0$ leads to a good fit of the
%experimental results of Ref.~\cite{Saito16}, which reported Ising superconductivity in
%ion-gated MoS$_2$.

Here, we only consider an $s$-wave order parameter in the singlet channel. 
The existence of singlet-triplet mixing of the order parameter has been
discussed~\cite{Moeckli20, IsingSC20_PRX, Cho21, Kuz21, Hamill21}. 
The presence of a triplet component and singlet-triplet mixing could possibly lead to new
and interesting properties of the mirage gaps.

The mirage gaps appear to be similar to the hybridization gaps in a two-band
superconductor~\cite{Komendova15}. 
%, which are also partially filled gaps with coherence peaks at higher energies. 
However, the underlying physics is quite different. The hybridization gaps are due to
single-quasiparticle scattering between two superconducting bands, while the mirage gaps
are the consequence of finite-energy pairing. 
%due to the interplay between noncollinear spins and the valley degree of freedom. 
Moreover, the mirage gaps are associated with triplet pairing correlations, while the
correlations in Ref.~\cite{Komendova15} are of singlet type. 
%Moreover, the mirage gap is tunable since it is created by magnetic field.  
Both the mirage gaps and hybridization gaps relate to the appearance of odd-frequency
pairing.

It is interesting to compare the finite-energy pairing state with the
Fulde-Ferrell-Larkin-Ovchinnikov (FFLO) state~\cite{FFLO,LOFF} 
in superconductors with large magnetic fields. 
For the FFLO state, due to the Zeeman splitting, two electrons at the Fermi surface with
the same energy can only pair with each other at the cost of a finite center-of-mass
momentum. In contrast to that, for the finite-energy pairing, the two electrons forming a
Cooper pair have the opposite momentum at the cost of different energies.

%Very recently, type-II Ising superconductivity in inversion-symmetric systems with
%spin-orbit locking at the $\Gamma$ point was proposed~\cite{IsingSC2_19PRL} and has been
%experimentally confirmed in few-layer stanene~\cite{IsingSC2_20Science} and PdTe$_2$
%films~\cite{IsingSC2_20NL}. In these systems, the signs of the ISOC fields are
%opposite for opposing orbitals. Our results can possibly be applied to type-II Ising
%superconductors as well.

Our findings should be experimentally accessible using tunneling
spectroscopy~\cite{Tinkham, Dvir18,Costanzo18, triplet-gap18} in an Ising superconductor with
moderate ISOC, such as MoS$_2$~\cite{Lu15, Saito16, Saito16_review}, by
applying an in-plane magnetic field. 
For Ising superconductors with large ISOC, such as NbSe$_2$~\cite{Xi16,
delaBarrera18, Cho21, Kuz21, Hamill21, Kang21}, WS$_2$~\cite{Lu18} and
TaS$_2$~\cite{delaBarrera18}, identifying the mirage gaps requires a relatively large
magnetic field according to Eq.~\eqref{delta}. This could be possibly realized using the
magnetic exchange field from a ferromagnetic substrate~\cite{Hamill21, Kang21}.
%, such as EuS~\cite{proximity_2, proximity_3}, CrI$_3$~\cite{proximity_1} or
%\red{CrBr$_3$~\cite{Hamill21, Kang21}.} 

To conclude, we have identified the emergence of finite-energy pairing correlations in an
Ising superconductor subject to an in-plane magnetic field. The accompanying mirage gaps
offer an experimental signature. 
%and can be identified using tunneling spectroscopy
The mirage gaps can also lead to equal-spin Andreev reflection at interfaces between an
Ising superconductor and a normal or ferromagnetic metal. A Josephson junction between two
Ising superconductors with noncollinear in-plane magnetic fields may host spin-polarized
Andreev bound states inside the mirage gaps that can be detected by spin-resolved
spectroscopy.  The concept of the mirage gap thus offers a new perspective on the
interplay between superconductivity and magnetism. 

\begin{acknowledgments}
{\it Acknowledgments.--} 
G.T. and C.B. acknowledge financial support from the Swiss National Science Foundation
(SNSF) and the NCCR Quantum Science and Technology.
W.B. acknowledges funding by the Deutsche Forschungsgemeinschaft (DFG, German Research
Foundation) -- Project-ID 443404566 - SPP 2244. 
\end{acknowledgments}

\bibliography{bib_IsingSC}{}

\clearpage

\begin{center}
  \large{\bf{Supplemental Material for ``Magnetic field-induced `mirage' gap in an Ising
  superconductor"}}
\end{center}

\section{Pairing-correlation function}
We consider the effective mean-field Hamiltonian near the $K$ or $K'$ valley with
\begin{equation}
  H_{\rm BdG} = 
  \begin{bmatrix}
    H_0({\bm k})  &  \Delta i\sigma_y  \\
    -\Delta i\sigma_y  &  -H_0^*(-{\bm k}) 
  \end{bmatrix} ,
\end{equation}
in the Nambu basis $(c_{{\bm k},\uparrow}, c_{{\bm k},\downarrow}, c_{-{\bm
k},\uparrow}^\dag, c_{-{\bm k},\downarrow}^\dag)$. The Hamiltonian $H_0$ is 
\begin{equation}
  H_0({\bm k}={\bm p}+s {\bm K})=\xi_{\bm p}\sigma_0 + s\beta_{\rm so} \sigma_z - {\bm B}
  \cdot \bm{\sigma} ,
\end{equation}
where $s{\bm K}$ is the momentum of the $K$ ($s=+$) or $K'$ ($s=-$) valley and ${\bm p}$
is the momentum deviation from ${\bm K}$ or $-{\bm K}$. Furthermore, $\xi_{\bm p}=|{\bm
p}|^2/(2m)-\mu$ is the dispersion measured form the chemical potential $\mu$. The Pauli
matrices $\sigma_x, \sigma_y$, and $\sigma_z$ act on the spin space and $\sigma_0$ is the
corresponding unit matrix.
In this work, we also define the Pauli matrices $\tau_1$, $\tau_2$ and $\tau_3$ acting on
the Nambu space with the corresponding unit matrix $\tau_0$. The Ising spin-orbit coupling
(ISOC) strength is denoted as $\beta_{\rm so}$. The Zeeman term is $-{\bm B} \cdot {\bm
\sigma}$ where the magnetic field is in the $x$-$z$ plane and parametrized as 
\begin{equation}
  {\bm B} = (B_x, 0, B_z) = B(\sin\theta, 0, \cos\theta) .        
\end{equation}
The magnetic field ${\bm B}$ absorbs the prefactor $g_L\mu_B$ with the Land\'{e} g-factor
$g_L$ and the Bohr magneton $\mu_B$. 

The pairing-correlation function $F({\bm k}, \varepsilon)$ with the form 
\begin{equation}
  F({\bm k}, \varepsilon) =\Delta\big[ F_0({\bm k}, \varepsilon)\sigma_0 + {\bm F}({\bm
  k}, \varepsilon)\cdot{\bm \sigma} \big] i\sigma_y ,
\end{equation}
can be obtained from the Gor'kov equation~\cite{noneqSC}
\begin{align}
  \begin{bmatrix}
     \varepsilon -H_0({\bm k})& -\Delta i\sigma_y\\
     \Delta i\sigma_y & \varepsilon +H^*_0(-{\bm k})\\
  \end{bmatrix}
  \begin{bmatrix}
    F({\bm k}, \varepsilon) \\ \bar{G}({\bm k}, \varepsilon)
  \end{bmatrix}
   =
  \begin{bmatrix}
    0  \\  1
  \end{bmatrix} . 
\end{align} 
One can obtain
\begin{align}
  F_0({\bm k}, \varepsilon) &= ({\varepsilon^2-\xi_{\bm p}^2-\Delta^2-\beta_{\rm
  so}^2}+B^2)/M({\bm k},\varepsilon) , \\
  F_x({\bm k}, \varepsilon) &= {-2\varepsilon B_x}/{M({\bm k},\varepsilon)}, \\
  F_y({\bm k}, \varepsilon) &= {-2i s\beta_{\rm so} B_x}/{M({\bm k},\varepsilon)}, \\
  F_z({\bm k}, \varepsilon) &= {-2(\varepsilon B_z -s\beta_{\rm so}\xi_{\bm p})}/{M({\bm
  k},\varepsilon)} ,
\end{align}
with
\begin{align}
  M({\bm k},\varepsilon) &= (\varepsilon^2-\xi_{\bm p}^2-\Delta^2-\beta_{\rm so}^2+B^2)^2
  \notag \\
 & -4(\varepsilon B_z-s\beta_{\rm so}\xi_{\bm p})^2 -4B_x^2(\varepsilon^2-\beta_{\rm
 so}^2). 
\end{align}

\section{Transformation of spin-quantization axis}
We consider a general spin-quantization orientation parametrized by the polar angle
($\theta$) and azimuthal angle ($\varphi$) with respect to the $z$-axis. The
transformation from the spin-quantization axis characterized by orientation
$(\theta,\varphi)$ to that by $z$-axis is given by 
\begin{equation}
  \begin{bmatrix}
    |{\uparrow}\rangle_{\theta,\varphi} \\ |{\downarrow}\rangle_{\theta,\varphi}
  \end{bmatrix}
  = 
  \begin{bmatrix}
    \cos(\theta/2) & \sin(\theta/2) \\
    -\sin(\theta/2) & \cos(\theta/2)
  \end{bmatrix}
  \begin{bmatrix}
    e^{-i\varphi/2}|{\uparrow}\rangle \\ e^{+i\varphi/2}|{\downarrow}\rangle
  \end{bmatrix} ,
\end{equation}
where $|{\uparrow}\rangle$ and $|{\downarrow}\rangle$ are spin vectors in the $z$-axis. 
This transformation rule applies to the Cooper pairs as well, for example,
\begin{align}
  (|{\uparrow}{\downarrow}\rangle -|{\downarrow}{\uparrow}\rangle)_{\theta,\varphi} =&
  |{\uparrow}{\downarrow}\rangle -|{\downarrow}{\uparrow}\rangle , \\
  (|{\uparrow}{\downarrow}\rangle +|{\downarrow}{\uparrow}\rangle)_{\theta,\varphi} =&
  -\sin\theta\big[e^{-i\varphi}|{\uparrow}{\uparrow}\rangle
    -e^{i\varphi}|{\downarrow}{\downarrow}\rangle\big] \notag \\
  &+\cos\theta\big[|{\uparrow}{\downarrow}\rangle +|{\downarrow}{\uparrow}\rangle\big] .
\end{align}
Specifically, the transformations of pairing forms from the $x$-axis with $\theta=\pi/2$ and
$\varphi=0$ to the $z$-axis are 
\begin{align}
  (|{\uparrow}{\downarrow}\rangle -|{\downarrow}{\uparrow}\rangle)_x =&
  |{\uparrow}{\downarrow}\rangle -|{\downarrow}{\uparrow}\rangle , \\
  (|{\uparrow}{\downarrow}\rangle +|{\downarrow}{\uparrow}\rangle)_x =&
  |{\downarrow}{\downarrow}\rangle -|{\uparrow}{\uparrow}\rangle , \\
  (|{\uparrow}{\uparrow}\rangle +|{\downarrow}{\downarrow}\rangle)_x =&
  |{\uparrow}{\uparrow}\rangle +|{\downarrow}{\downarrow}\rangle , \\
  (|{\uparrow}{\uparrow}\rangle -|{\downarrow}{\downarrow}\rangle)_x =&
  |{\uparrow}{\downarrow}\rangle +|{\downarrow}{\uparrow}\rangle .
\end{align}

\section{Quasiclassical Green's function}
The quasiclassical formalism concentrates on the phenomena close to the Fermi
surface~\cite{Eilenberger1968, LO1969, Belzig99, noneqSC}, it can be applied to the
situation where both the superconducting gap and the ISOC are much smaller than the
Fermi energy. The structure of the quasiclassical Green's function is \cite{noneqSC,
Eschrig15}
\begin{equation}
  \hat{g}(\hat{\bm k}, \varepsilon) =
  \begin{bmatrix}
    g_0\sigma_0 + \bm{g}\cdot \bm{\sigma} & (f_0\sigma_0 + \bm{f}\cdot \bm{\sigma}) 
    i\sigma_y \\
    (\bar{f}_0\sigma_0 + \bar{\bm{f}} \cdot \bm{\sigma}^*)i\sigma_y  & \bar{g}_0
    \sigma_0+\bar{\bm{g}}\cdot {\bm \sigma}^* 
  \end{bmatrix} ,
\end{equation}
where $\hat{\bm k}$ denotes the direction of momentum ${\bm k}$ and the bar operation
is defined as $\bar{q}(\hat{\bm k}, \varepsilon) = q(-\hat{{\bm k}}, -\varepsilon^*)^*$
with $q\in \{ g_0, f_0, {\bm g}, {\bm f}, \Delta, {\bm \Delta}, {\bm \nu} \}$.
We introduce the notation ${\bm g}_{\pm}=({\bm g}\pm \bar{\bm g})/2$. The anomalous
Green's functions $f_0$ and $\bm{f}$ characterize the singlet and triplet pairings,
respectively. 
The requirement that ${\rm Tr}\big( \hat{g} \big) =0$ leads to $\bar{g}_0 = -g_0$. The
normalization condition $\hat{g} \hat{g} =\sigma_0\tau_0$ gives
\begin{align}
  &g_0^2 + {\bm g}_+^2 + {\bm g}_-^2 - f_0\bar{f}_0 +{\bm f}\cdot \bar{\bm f} =1,
  \label{norm} \\
  &2g_0 {\bm g}_+ = \bar{f}_0 {\bm f}- f_0 \bar{{\bm f}}, \label{g+} \\
  &2g_0 {\bm g}_- = i \bar{\bm f}\times {\bm f}. \label{g-}
\end{align}
with ${\bm g}_{\pm }^2 = {\bm g}_{\pm, x}^2 +{\bm g}_{\pm, y}^2 + {\bm g}_{\pm, z}^2$.

The quasiclassical Green's function $\hat{g}(\hat{\bm k}, \varepsilon)$ obeys the
Eilenberger equation~\cite{Eilenberger1968, noneqSC},
\begin{equation} \label{Eilen}
  \big[\varepsilon \sigma_0\tau_3 -\hat{\Delta}-\hat{\nu} -\hat{\Sigma}(\varepsilon),
  \hat{g} \big] +i\hbar {\bm v}_F \cdot \nabla_{\bm R} \hat{g} =0. 
\end{equation}
Here, the order parameter term $\hat{\Delta}$ is explicitly written as 
\begin{equation}
\hat{\Delta} = 
  \begin{bmatrix}
    & (\Delta\sigma_0 +{\bm \Delta}\cdot {\bm \sigma})i\sigma_y \\
    (\bar{\Delta}\sigma_0 +\bar{\bm \Delta}\cdot {\bm \sigma}^*)i\sigma_y &  
  \end{bmatrix} .
\end{equation}
The ISOC and Zeeman fields are included in $\hat{\nu}$ with
\begin{equation}
  \hat{\nu} =
  \begin{bmatrix}
    {\bm \nu}\cdot{\bm \sigma} & \\  & \bar{\bm \nu}\cdot {\bm \sigma^*}
  \end{bmatrix} 
  = s\beta_{\rm so}\sigma_z \tau_3 -
  \begin{bmatrix}
    {\bm B}\cdot {\bm \sigma} &  \\
     & {\bm B}\cdot {\bm \sigma}^*
  \end{bmatrix} .
\end{equation}
%which implies that
%\begin{align}
%  \nu_x &= -B_x , \qquad  \nu_z = +s\beta_{\rm so}-B_z , \\
%  \bar{\nu}_x &= -B_x , \qquad  \bar{\nu}_z = -s\beta_{\rm so}-B_z .
%\end{align}
By introducing the notation ${\bm \nu}_{\pm}=({\bm \nu}\pm \bar{\bm \nu})/2$, we have
\begin{equation}
  \hat{\nu} =
  \begin{bmatrix}
    ({\bm \nu}_+ + {\bm \nu}_-)\cdot {\bm \sigma} & \\
      & ({\bm \nu}_+ - {\bm \nu}_-)\cdot {\bm \sigma}^*  
  \end{bmatrix} ,
\end{equation}
with
\begin{align}
  \nu_{+,x} &= -B_x,  &&\nu_{+,z} = -B_z ,\\
  \nu_{-,x} &= 0,  &&\nu_{-,z} = s\beta_{\rm so} .
\end{align}
The nonmagnetic impurities are considered within the self-consistent Born approximation
with
\begin{equation}
  \hat{\Sigma}(\varepsilon) = -i\Gamma\langle \hat{g}(\hat{\bm k}, \varepsilon)\rangle,
\end{equation}
where $\Gamma$ is the intervalley impurity scattering strength and $\langle \cdots
\rangle$ denotes averaging over the whole Fermi momentum direction.  
%$\Gamma = 1/(2\tau_{\rm imp})$

\section{Clean limit}
For a homogeneous system ($\nabla_{\bm R} \hat{g}=0$) in the clean limit ($\Gamma=0$), the
Eilenberger equation, Eq.~\eqref{Eilen}, is reduced to
\begin{equation} \label{Eilen_clean}
  \big[\varepsilon \hat{\tau}_3 -\hat{\Delta}-\hat{\nu}, \hat{g} \big] =0. 
\end{equation}
The off-diagonal terms in the Nambu space produce
\begin{align}
  &\varepsilon f_0 -{\bm \nu}_+\cdot {\bm f} +\Delta g_0 +{\bm \Delta}\cdot{\bm g}_+=0,
  \label{g1} \\
  &\varepsilon {\bm f}-f_0{\bm \nu}_+-i{\bm \nu}_-\times {\bm f} +g_0 {\bm \Delta}
  +\Delta {\bm g}_+ + i{\bm g}_-\times {\bm \Delta}=0, \label{g2} \\
  &-\varepsilon \bar{f}_0 -{\bm \nu}_+\cdot \bar{\bm f} -\bar{\Delta} g_0 +\bar{\bm
  \Delta}\cdot{\bm g}_+=0, \label{g3} \\
  &-\varepsilon \bar{\bm f}-\bar{f}_0{\bm \nu}_+-i{\bm \nu}_-\times \bar{\bm f} -g_0 \bar{\bm
  \Delta} +\bar{\Delta} {\bm g}_+ + i{\bm g}_-\times \bar{\bm \Delta}=0. \label{g4}
\end{align}
In combination with Eqs.~\eqref{norm}--\eqref{g-}, one can obtain all the components of the
quasiclassical Green's function $\hat{g}$.  
Considering a Hamiltonian with only an $s$-wave singlet pairing (${\bm \Delta} =0$),
Eqs.~\eqref{g1}--\eqref{g4} are reduced to
\begin{align}
  &\varepsilon f_0 -{\bm \nu}_+\cdot {\bm f} +\Delta e^{i\phi} g_0 =0, \label{f0} \\
  &\varepsilon {\bm f} -f_0{\bm \nu}_+-i{\bm \nu}_-\times {\bm f} +\Delta e^{i\phi} {\bm
  g}_+ =0, \label{ff} \\
  &\varepsilon \bar{f}_0 +{\bm \nu}_+\cdot \bar{\bm f} +\Delta e^{-i\phi} g_0 =0,
  \label{bf0} \\
  &\varepsilon \bar{\bm f} +\bar{f}_0{\bm \nu}_+ +i{\bm \nu}_-\times \bar{\bm f} 
  -\Delta e^{-i\phi} {\bm g}_+ =0, \label{bff}
\end{align}
where $\phi$ is the superconducting phase. Equations~\eqref{f0} and \eqref{ff} can be written
explicitly as
\begin{equation} \label{f0ff}
  \begin{bmatrix}
    \varepsilon & B_x & 0 & B_z \\
    B_x & \varepsilon & is\beta_{\rm so} & 0 \\
    0 & -is\beta_{\rm so} & \varepsilon & 0 \\
    B_z & 0 & 0 & \varepsilon  
  \end{bmatrix}
  \begin{bmatrix}
    f_0 \\ f_x \\ f_y \\ f_z
  \end{bmatrix} 
  + \Delta e^{i\phi}
  \begin{bmatrix}
    g_0 \\ g_{+,x} \\ g_{+,y} \\ g_{+,z} \\
  \end{bmatrix} 
  =0 .
\end{equation}

We now try to obtain some relations between ($f_0$, ${\bm f}$) and ($\bar{f}_0$, $\bar{\bm
f}$). By combining Eqs.~\eqref{f0} and \eqref{bf0}, we have
\begin{equation}
  e^{-i\phi}(\varepsilon f_0 -{\bm \nu}_+\cdot {\bm f}) = e^{i\phi}(\varepsilon \bar{f}_0
  +{\bm \nu}_+\cdot \bar{\bm f}) .
\end{equation}
By combining Eqs.~\eqref{ff} and \eqref{bff}, we have
\begin{equation}
  e^{-i\phi}(\varepsilon {\bm f} - f_0{\bm \nu}_+ -i{\bm \nu}_-\times {\bm f}) =
  -e^{i\phi}(\varepsilon \bar{\bm f} -\bar{f}_0{\bm \nu}_+ +i{\bm \nu}_-\times \bar{\bm
  f}).
\end{equation}
The above two equations can be explicitly written as 
\begin{align}
  &\varepsilon (f_0-\bar{f}_0)+B_x(f_x+\bar{f}_x)+B_z(f_z+\bar{f}_z)=0,
  \label{ef0} \\
  &\varepsilon (f_x+\bar{f}_x)+B_x(f_0-\bar{f}_0)+is\beta_{\rm so}(f_y-\bar{f}_y)=0 ,
  \label{efx} \\
  &\varepsilon (f_y+\bar{f}_y)-is\beta_{\rm so}(f_x-\bar{f}_x)=0 ,
  \label{efy} \\
  &\varepsilon (f_z+\bar{f}_z)+B_z(f_0-\bar{f}_0)=0 ,
  \label{efz}
\end{align}
where the phase factors $e^{-i\phi}$ and $e^{i\phi}$ are absorbed into ($f_0$, ${\bm f}$)
and ($\bar{f}_0$, $\bar{\bm f}$), respectively. From Eqs.~\eqref{ef0}--\eqref{efz}, we have
\begin{equation} \label{0xyz}
  f_0 = \bar{f}_0, \quad f_x =-\bar{f}_x, \quad f_y = \bar{f}_y, \quad f_z =-\bar{f}_z,
\end{equation}
and
\begin{equation}
  \varepsilon f_y = is\beta_{\rm so} f_x .
\end{equation}
Equation~\eqref{0xyz} results in $g_{\pm ,y} = 0$ from Eqs.~\eqref{g+} and \eqref{g-}.
Then Eqs.~\eqref{norm}--\eqref{g-} can be written explicitly as
\begin{align}
  & g_0^2 +g_{+,x}^2 +g_{+,z}^2 +g_{-,x}^2 +g_{-,z}^2 -f_0^2 -f_x^2 +f_y^2 -f_z^2 =1,
  \label{norm2} \\
  & \quad g_0 g_{+,x} = f_0 f_x,    \quad \quad g_0 g_{+,z} = f_0 f_z, \label{g+2} \\
  & \quad g_0 g_{-,x} = +i f_y f_z, \quad       g_0 g_{-,z} = -i f_x f_y. \label{g-2}
\end{align}
In the following, we obtain the corresponding components of the Green's function $\hat{g}$
from Eq.~\eqref{f0ff} and Eqs.~\eqref{0xyz}--\eqref{g-2} for three different cases.

For the case of a magnetic field with an arbitrary direction, we can express
\begin{equation} \label{fxfy2}
  f_x = a\ \varepsilon B_x , \quad f_y = a\ is\beta_{\rm so} B_x ,
  \quad f_z = b\ \varepsilon B_z ,
\end{equation}
where $a$ and $b$ are to be fixed by the normalization condition, Eq.~\eqref{norm2}. From
Eqs.~\eqref{f0ff} and \eqref{g+2},
we can obtain
\begin{equation}
  (a-b) f_0 = -ab\beta_{\rm so}^2 ,
\end{equation}
where $a$ and $b$ satisfy
\begin{align*}
  & a^2 b\beta_{\rm so}^4-a(a-b)(b\varepsilon^2+aB_x^2+bB_z^2-b\Delta^2)\beta_{\rm so}^2
  \\
  + & (a-b)^2 \varepsilon^2 (aB_x^2 + bB_z^2) =0 .
\end{align*}
%By letting $b-a=\beta_{\rm so}^2 c$, we have
%\begin{equation}
%  f_z = (\beta_{\rm so}^2 c +a)\varepsilon B_z , \quad
%  f_0 = \frac{a^2}{c} + a\beta_{\rm so}^2 ,
%\end{equation}
%with
%\begin{align} \label{ac}
%  & a^3 +a^2c(\varepsilon^2 +B_x^2 +B_z^2 +\beta_{\rm so}^2-\Delta^2) \notag \\
%  + &ac^2\big[ \varepsilon^2(B_x^2+B_z^2+\beta_{\rm so}^2)+\beta_{\rm
%  so}^2(B_z^2-\Delta^2) \big] \notag \\
%  + &c^3\varepsilon^2\beta_{\rm so}^2 B_z^2 =0.
%\end{align}
Then $g_0$ can be obtained from the first relation in Eq.~\eqref{f0ff}, which is
\begin{equation}
  \varepsilon f_0 +B_x f_x +B_z f_z + \Delta g_0 =0. 
\end{equation}
The terms $g_{+,x}$ and $g_{+,z}$ can be obtained from Eq.~\eqref{g+2}, and $g_{-,x}$ and
$g_{-,z}$ from Eq.~\eqref{g-2}. 

For the case with an out-of-plane magnetic field $B_z$, both the $x$- and $y$-components
are decoupled so that $f_x=f_y=g_{+,x}=g_{+,y}=0$, which leads to ${\bm g}_-=0$ from
Eq.~\eqref{g-2}. From Eqs.~\eqref{f0ff} and \eqref{g+2} and the normalization condition,
Eq.~\eqref{norm2}, we can directly get the Green's functions.
By defining $f_{\uparrow(\downarrow)}=f_0\pm f_z$ and $g_{\uparrow(\downarrow)}=g_0\pm
g_{+,z}$, so that $f_{0(z)}=(f_{\uparrow}\pm f_{\downarrow})/2$ and
$g_{0(+,z)}=(g_{\uparrow}\pm g_{\downarrow})/2$, we have
\begin{equation}
  g_{\uparrow(\downarrow)} = \frac{\varepsilon\pm B_z}{\sqrt{(\varepsilon \pm
  B_z)^2-\Delta^2}}, \quad
  f_{\uparrow(\downarrow)} = \frac{-\Delta}{\sqrt{(\varepsilon \pm B_z)^2-\Delta^2}}.
\end{equation}

For the case of an in-plane magnetic field $B_x$, one can see from Eq.~\eqref{f0ff} that
both $f_z$ and $g_{+,z}$ are decoupled from the other components so that $f_z=g_{+,z}=0$.
We have
\begin{equation} \label{fxfy}
  f_x = a\ \varepsilon B_x , \qquad f_y = a\ is\beta_{\rm so} B_x .
\end{equation}
From Eqs.~\eqref{f0ff} and \eqref{g+2}, the expressions of $g_0$ and $f_0$ are obtained as
\begin{align}
  g_0 &= a\varepsilon \ c /(2\Delta), \\
  f_0 &= -a\ (B_x^2 + c/2), 
\end{align}
%\begin{align}
%  f_0 &= a\ \frac{-\varepsilon^2+\beta_{\rm so}^2-B_x^2+\Delta^2 +u}{2},\\
%  g_0 &= a\varepsilon \ \frac{\varepsilon^2-\beta_{\rm so}^2-B_x^2-\Delta^2 -u}{2\Delta},
%  \\
%  g_{+,x} &= a B_x \ \frac{-\varepsilon^2+\beta_{\rm so}^2+B_x^2-\Delta^2 -u}{2\Delta}, 
%\end{align}
where
\begin{equation}
  c = \varepsilon^2-\beta_{\rm so}^2-B_x^2-\Delta^2 -u
\end{equation}
with
\begin{equation} \label{u}
  u = \pm\sqrt{(\varepsilon^2-\beta_{\rm so}^2-B_x^2-\Delta^2)^2-4B_x^2\Delta^2} .
\end{equation}
The terms $g_{+,x}$ and $g_{-,z}$ can be obtained from Eqs.~\eqref{g+2} and \eqref{g-2},
respectively. 
%\begin{equation}
%  g_{-,z} = a s\beta_{\rm so} \ \frac{\varepsilon^2-\beta_{\rm so}^2 -B_x^2-\Delta^2
%  +u}{2\Delta} .
%\end{equation}
The coefficient $a$ can be fixed by the normalization condition 
\begin{equation}
  g_0^2 + g_{+,x}^2 + g_{-,z}^2 - f_0^2 -f_x^2 +f_y^2 =1 ,
\end{equation}
so that
\begin{equation}
  a^2 =\frac{4c^2 \Delta^2}{[4\Delta^2 B_x^2 (B_x^2+\beta_{\rm so}^2+c)
  +c^2(\Delta^2-\varepsilon^2)] (4\Delta^2B_x^2-c^2)} ,
\end{equation}
which can be further simplified as
\begin{equation}
  a^2 =\frac{c \Delta^2}{u^2 [ c(\varepsilon^2-\Delta^2)-2\Delta^2B_x^2 ]}.
\end{equation}
%\begin{equation}
%  a =\frac{\sqrt{c} \Delta}{u \sqrt{ c(\varepsilon^2-\Delta^2)-2\Delta^2B_x^2 }}.
%\end{equation}
Since $a$ is even in energy and does not depend on the valley index $s$, $f_x$ is even in
valley index and odd in energy, while $f_y$ is odd in valley index and even in energy.

\begin{figure*}
\centering
\includegraphics[width=5.5in]{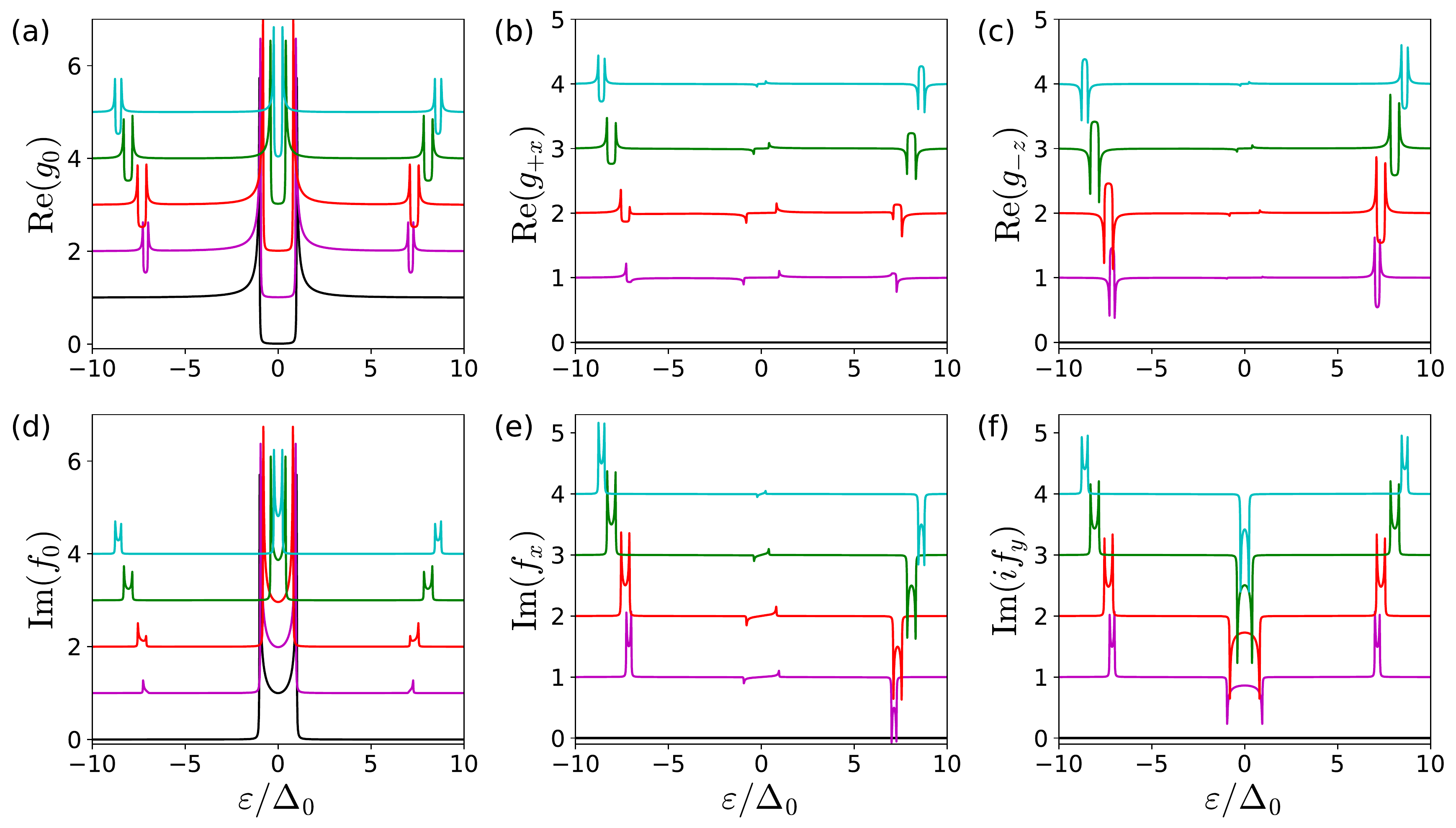}
\caption{Various components of the Green's function at different in-plane magnetic fields
  (from bottom to top: $B_x=0$, $\Delta_0$, $2\Delta_0$, $4\Delta_0$, $5\Delta_0$) in the
  clean limit. All curves for finite $B_x$ have been offset vertically for better
  visibility. Here, $\beta_{\rm so}=7\Delta_0$ and $T=0.1T_{c0}$. }
\label{fig_gf}
\end{figure*}

\begin{figure}
\centering
\includegraphics[width=\columnwidth]{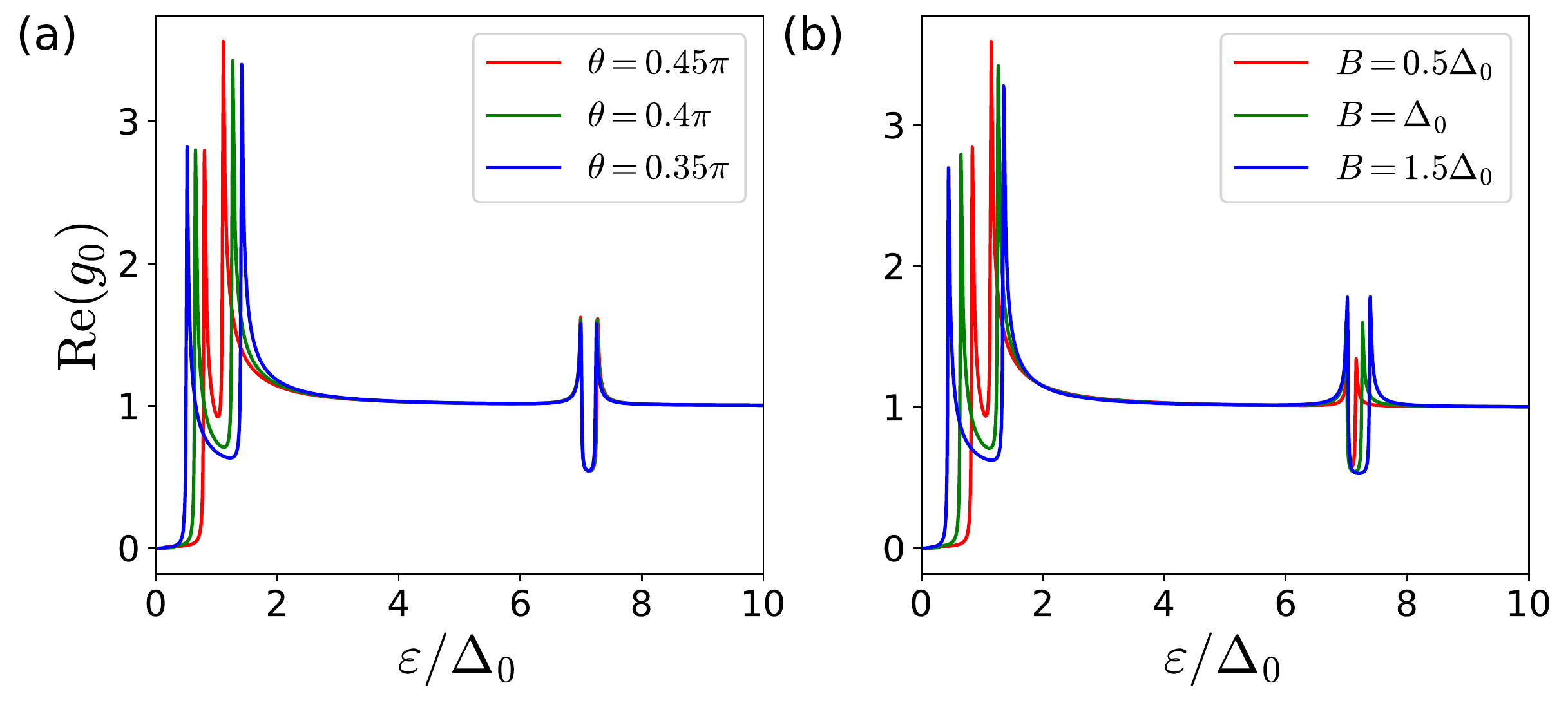}
\caption{Density of states ${\rm Re}(g_0)$ (a) at different $\theta$ with $B=\Delta_0$ and
(b) at different $B$ with $\theta=0.4\pi$ in the clean limit. Here, $\beta_{\rm
so}=7\Delta_0$ and $T=0.1T_c$. }
\label{fig_gf_B}
\end{figure}

In Fig.~\ref{fig_gf}, we show various components of the Green's function at different
in-plane magnetic fields in the clean limit. 
In Fig.~\ref{fig_gf_B}(a), we show the effects of the magnetic field direction on the DOS
with the magnitudes of the magnetic field fixed. The out-of-plane component causes the
spin splitting near the superconducting gap and the mirage gaps are not influenced. The
effects of the magnetic field magnitude with a fixed direction are shown in
Fig.~\ref{fig_gf_B}(b) where we can see that the mirage gap depends on the magnetic field
magnitude. 

\section{nonmagnetic impurity effect}
The Eilenberger equation for a homogeneous system with impurities reads
\begin{equation} 
  \big[\varepsilon \hat{\tau}_3 -\hat{\Delta}-\hat{\nu} -\hat{\Sigma}(\varepsilon),
  \hat{g} \big] =0,
\end{equation}
with $\hat{\Sigma}(\varepsilon) = -i\Gamma\langle \hat{g}(\hat{\bm k}, \varepsilon)
\rangle$ where $\Gamma$ is the nonmagnetic intervalley impurity scattering rate. 
Considering an in-plane magnetic field, we have $f_z=0$. Due to the fact that $f_y$ and
$g_{-,z}$ are odd with respect to the valley index $s$, we have $\langle f_y \rangle
=\langle g_{-,z}\rangle=0$. 
Since the other components of the quasi-classical Green's function are independent of the
momentum, the average notation $\langle \cdots \rangle$ can be omitted in
$\hat{\Sigma}(\varepsilon) = -i\Gamma\langle \hat{g}(\hat{\bm k}, \varepsilon)\rangle$.
The Eilenberger equation can be recast into
\begin{equation} 
  \big[\tilde{\varepsilon} \sigma_0\tau_3 -\hat{\tilde{\Delta}}-\hat{\tilde{\nu}}, \hat{g}
  \big] =0 ,
\end{equation}
with $\tilde{\varepsilon}=\varepsilon +i\Gamma g_0$. The effective superconducting gap is
\begin{equation}
  \hat{\tilde{\Delta}} = 
  \begin{bmatrix}
    & (\tilde{\Delta}\sigma_0 +\tilde{\bm \Delta}\cdot {\bm \sigma})i\sigma_y \\
    (\bar{\tilde{\Delta}}\sigma_0 +\bar{\tilde{\bm \Delta}}\cdot {\bm \sigma}^*)i\sigma_y &  
  \end{bmatrix} ,
\end{equation}
with $\tilde{\Delta}=\Delta e^{i\phi} -i\Gamma f_0$, $\tilde{\Delta}_x= -i\Gamma f_x$,
$\bar{\tilde{\Delta}}=\Delta e^{-i\phi} -i\Gamma \bar{f}_0$, $\bar{\tilde{\Delta}}_x=
-i\Gamma \bar{f}_x$, and $\tilde{\Delta}_y =\tilde{\Delta}_z =\bar{\tilde{\Delta}}_y
=\bar{\tilde{\Delta}}_z =0$.  We also have
\begin{equation}
  \hat{\tilde{\nu}} =
  \begin{bmatrix}
    (\tilde{\bm \nu}_+ + \tilde{\bm \nu}_-)\cdot {\bm \sigma} & \\
      & (\tilde{\bm \nu}_+ - \tilde{\bm \nu}_-)\cdot {\bm \sigma}^*  
  \end{bmatrix} ,
\end{equation}
with
\begin{align}
  \tilde{\nu}_{+,x} &= -\tilde{B}_x,  &&\tilde{\nu}_{+,z} = 0 ,\\
  \tilde{\nu}_{-,x} &= 0,  &&\tilde{\nu}_{-,z} = s\beta_{\rm so} ,
\end{align}
where $\tilde{B}_x = B_x +i\Gamma g_{+,x}$. 

Equations~\eqref{g1}-\eqref{g4} become
\begin{align*}
  &\tilde{\varepsilon} f_0 -\tilde{\bm \nu}_+\cdot {\bm f} +\tilde{\Delta} g_0
    +\tilde{\bm \Delta}\cdot{\bm g}_+=0, \\
  &\tilde{\varepsilon} {\bm f}-f_0\tilde{\bm \nu}_+-i\tilde{\bm \nu}_-\times {\bm f} +g_0
  \tilde{\bm \Delta} +\tilde{\Delta} {\bm g}_+ =0, \\
  &\tilde{\varepsilon} \bar{f}_0 +\tilde{\bm \nu}_+\cdot \bar{\bm f} +\bar{\tilde{\Delta}}
  g_0 -\bar{\tilde{\bm \Delta}}\cdot{\bm g}_+=0,  \\
  &\tilde{\varepsilon} \bar{\bm f} +\bar{f}_0 \tilde{\bm \nu}_+ +i\tilde{\bm \nu}_-\times
  \bar{\bm f} +g_0 \bar{\tilde{\bm \Delta}} -\bar{\tilde{\Delta}} {\bm g}_+ =0 ,
\end{align*}
which can be explicitly written as
\begin{equation} \label{f0ff_i}
  \begin{bmatrix}
    \tilde{\varepsilon} & \tilde{B}_x & 0 \\
    \tilde{B}_x & \tilde{\varepsilon} & is\beta_{\rm so} \\
    0 & -is\beta_{\rm so} & \tilde{\varepsilon }
  \end{bmatrix}
  \begin{bmatrix}
    f_0 \\ f_x \\ f_y
  \end{bmatrix} 
  +
  \begin{bmatrix}
    \tilde{\Delta} & \tilde{\Delta}_x & 0 \\
    \tilde{\Delta}_x & \tilde{\Delta} & 0 \\
    0 & 0 & \tilde{\Delta}
  \end{bmatrix}
  \begin{bmatrix}
    g_0 \\ g_{+,x} \\ 0
  \end{bmatrix} 
  =0 ,
\end{equation}
and
\begin{equation} 
  \begin{bmatrix}
    \tilde{\varepsilon} & -\tilde{B}_x & 0 \\
    \tilde{B}_x & -\tilde{\varepsilon} & is\beta_{\rm so} \\
    0 & -is\beta_{\rm so} & -\tilde{\varepsilon}
  \end{bmatrix}
  \begin{bmatrix}
    \bar{f}_0 \\ \bar{f}_x \\ \bar{f}_y
  \end{bmatrix} 
  +
  \begin{bmatrix}
    \bar{\tilde{\Delta}} & -\bar{\tilde{\Delta}}_x & 0 \\
    -\bar{\tilde{\Delta}}_x & \bar{\tilde{\Delta}} & 0 \\
    0 & 0 & \bar{\tilde{\Delta}}
  \end{bmatrix}
  \begin{bmatrix}
    g_0 \\ g_{+,x} \\ 0
  \end{bmatrix}
  =0 .
\end{equation}
From these equations, we still have the relations
\begin{equation}
  f_0 = \bar{f}_0, \quad f_x =-\bar{f}_x, \quad f_y = \bar{f}_y,
\end{equation}
and
\begin{equation}
  \tilde{\varepsilon} f_y = is\beta_{\rm so} f_x .
\end{equation}
Similarly to Eq.~\eqref{fxfy}, we can assume $f_x$ and $f_y$ as
\begin{equation}
  f_x = a\ \tilde{\varepsilon} B_x , \qquad f_y = a\ is\beta_{\rm so} B_x ,
\end{equation}
where $a$ is fixed by the normalization condition,
\begin{equation} 
  g_0^2 + g_{+,x}^2 + g_{-,z}^2 - f_0^2 -f_x^2 +f_y^2 =1 .
\end{equation}
From Eqs.~\eqref{f0ff_i} and \eqref{g+}, which is $g_0 g_{+,x} =f_0 f_x$, we have
\begin{equation} \label{g0}
  g_0 = a \varepsilon \ \frac{\varepsilon^2-\beta_{\rm so}^2-B_x^2-\Delta^2-2i\Gamma
  aB_x^2 \Delta- u}{2\Delta(1-\Gamma^2 a^2 B_x^2)-2i\Gamma a
  (\varepsilon^2-B_x^2-\Delta^2)} \equiv a \varepsilon \ \frac{c}{d} , 
\end{equation}
where 
\begin{equation} 
  u=\pm\sqrt{(\varepsilon^2-\beta_{\rm so}^2-B_x^2-\Delta^2)^2-4B_x^2(\Delta^2-i\Gamma
  a\Delta \beta_{\rm so}^2)}.
\end{equation}
Using the normalization condition, we have 
\begin{align} \label{nonlinear}
  & a^2 \Big\{ \big[ B_x^2 (d+ i\Gamma c a) +c\Delta \big]^2 -c^2\varepsilon^2
  +B_x^2 \beta_{\rm so}^2 d^2 \Big\} \times \notag \\
  & \big[B_x^2(d+i\Gamma c a)^2-c^2\big] -c^2 d^2 =0 .
\end{align}
Our task now is to get $a$ from Eq.~\eqref{nonlinear}.

\begin{figure}
\centering
\includegraphics[width=2.3in]{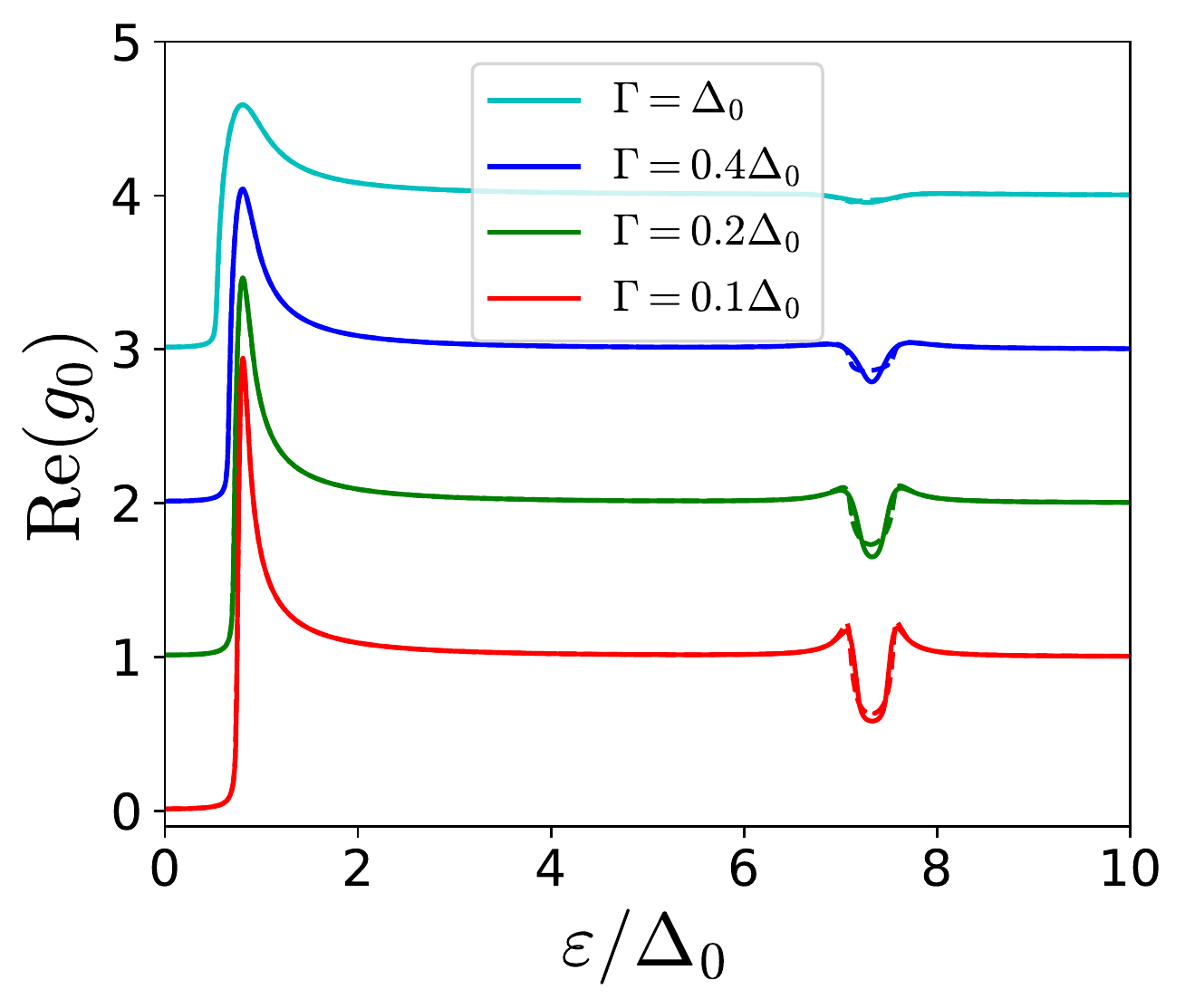}
\caption{Comparison of the density of states between the exact (solid lines) and
approximate results (dashed lines) for different impurity scattering strengths.
Here, $\beta_{\rm so}=7\Delta_0$, $B_x=2\Delta_0$ and $T=0.1T_{c0}$. }
\label{fig_compare}
\end{figure}

Since $a \sim {\Delta/(\varepsilon\beta_{\rm so}^2)}$ and we are mainly interested in the
DOS for $|\varepsilon| \gtrsim \Delta$, the quantity $\Gamma a B_x$ can be considered to
be small under the condition $\Gamma B_x \ll \beta_{\rm so}^2$.
Considering a small intervalley scattering strength $\Gamma$ with $\Gamma B_x \ll
\beta_{\rm so}^2$, one has
\begin{equation} \label{g0_approx}
  g_0 \approx a \varepsilon \ \frac{(\varepsilon^2-\beta_{\rm
  so}^2-B_x^2-\Delta^2)-u}{2\Delta-2i\Gamma a (\varepsilon^2-B_x^2-\Delta^2)} ,
\end{equation}
with $u \approx \pm \sqrt{(\varepsilon^2-\beta_{\rm
so}^2-B_x^2-\Delta^2)^2-4B_x^2\Delta^2}$.
Equation~\eqref{nonlinear} reduces to a polynomial equation of degree six with respect to
$a$. To obtain the coefficients of the polynomial using {\it Mathematica}, we write $d
+i\Gamma c a =2\Delta -i\Gamma x a$ with $x=\varepsilon^2 -B_x^2 -\Delta^2 +\beta_{\rm
so}^2+u$ and $d =2\Delta -i\Gamma y a$ with $y=2(\varepsilon^2 -B_x^2 -\Delta^2)$, so that
\begin{align} 
  & \Gamma^4 B_x^4 x^2(B_x^2 x^2 + \beta_{\rm so}^2 y^2)a^6 
  \notag \\
  +& 2iB_x^4\Gamma^3\Delta x \big[(4B_x^2 +c)x^2 + 2y(x+y)\beta_{\rm so}^2 \big]a^5 
  \notag \\
  +& B_x^2\Gamma^2 \big[c^2x^2(\varepsilon^2+B_x^2-\Delta^2)+\beta_{\rm so}^2 c^2 y^2 
  \notag \\
   & -12B_x^2 \Delta^2 x^2(2B_x^2+c)-4B_x^2\Delta^2\beta_{\rm so}^2(x^2+4xy+y^2) \big]a^4 
  \notag \\
  -& 2i\Gamma B_x^2\Delta \big[ 4B_x^2\Delta^2x(4B_x^2+3c) + 8B_x^2\beta_{\rm
    so}^2\Delta^2 (x+y) \notag \\
   & -c^2 x(2\varepsilon^2 +2B_x^2 -2\Delta^2 +c) -2c^2\beta_{\rm so}^2 y \big] a^3 
  \notag \\
  +& \big\{ \Gamma^2c^2y^2 +[4\Delta^2 B_x^2 (B_x^2+\beta_{\rm so}^2+c)
    +c^2(\Delta^2-\varepsilon^2)] \notag \\
   & (4\Delta^2B_x^2-c^2) \big\} a^2 +4i\Gamma c^2 y\Delta a -4c^2 \Delta^2 =0. 
   \label{sextic}
\end{align}
An exact solution for the coefficient $a$ can be numerically obtained
using {\it NLsolve.jl} package in {\it Julia} where the initial value can be assigned
using the approximated solution in the above equation. 
The terms $f_0$, $g_{+,x}$ and $g_{-,z}$ can be obtained from 
\begin{align}
  &\varepsilon f_0 + a\varepsilon B_x^2 + (i\Gamma a B_x^2 + \Delta) g_0 =0, \\
  &g_0 g_{+,x} = f_0 f_x, \\
  &g_0 g_{-,z} = -i f_x f_y , 
\end{align}
respectively.
The comparison between the approximate solution using Eq.~\eqref{sextic} and the exact
result for the density of states is shown in Fig.~\ref{fig_compare}. 

\section{Gap equation}
\begin{figure}
\centering
\includegraphics[width=\columnwidth]{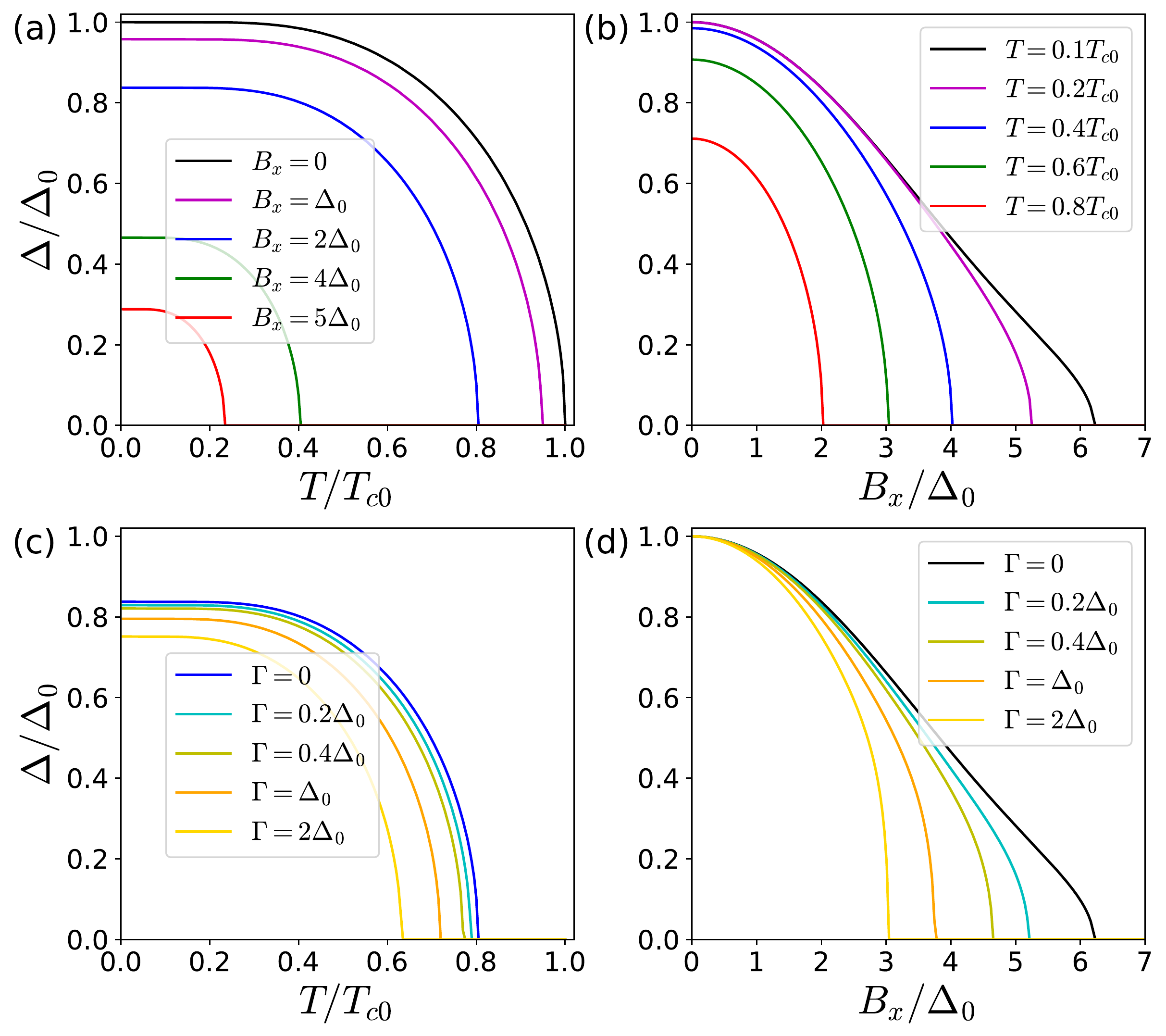}
\caption{(a) Superconducting gap $\Delta$ versus temperature $T$ at different magnetic
  fields $B_x$. (b) Gap $\Delta$ versus $B_x$ at different temperatures. Panels (a) and
  (b) are in the clean limit. 
  (c) Gap $\Delta$ versus $T$ at different intervalley scattering strengths $\Gamma$
  with $B_x=2\Delta_0$. (d) Gap $\Delta$ versus $B_x$ at different $\Gamma$ with
  $T=0.1T_c$. Here, $\beta_{\rm so}=7\Delta_0$.}
\label{fig_gap_Bx}
\end{figure}

The superconducting gap is determined from the self-consistent gap equation
\begin{equation}
  \frac{\Delta}{\lambda} = -2i\pi T \sum_{n=0}^{N_c} f_0(i\omega_n),
\end{equation}
where $\lambda$ is the dimensionless coupling constant and $\omega_n=(2n+1)\pi k_B T$
the Matsubara frequency. The summation cutoff $N_c$ is determined by the cutoff frequency
$\Omega_c$ with $N_c=\lfloor\hbar\Omega_c/(2\pi k_B T)\rfloor$. The term of
$f_0(i\omega_n)$ is obtained from $f(\varepsilon)$ by replacing $\varepsilon$ with
$i\omega_n$. In the clean limit without a magnetic field, we have $\hbar\Omega_c =
\Delta_0 \sinh(1/\lambda)$~\cite{Tinkham}, where $\Delta_0$ is the zero-temperature gap in
the absence of a magnetic field. 
The behaviors of the superconducting gap $\Delta$ are shown in Fig.~\ref{fig_gap_Bx}.

\begin{figure}
\centering
\includegraphics[width=2.3in]{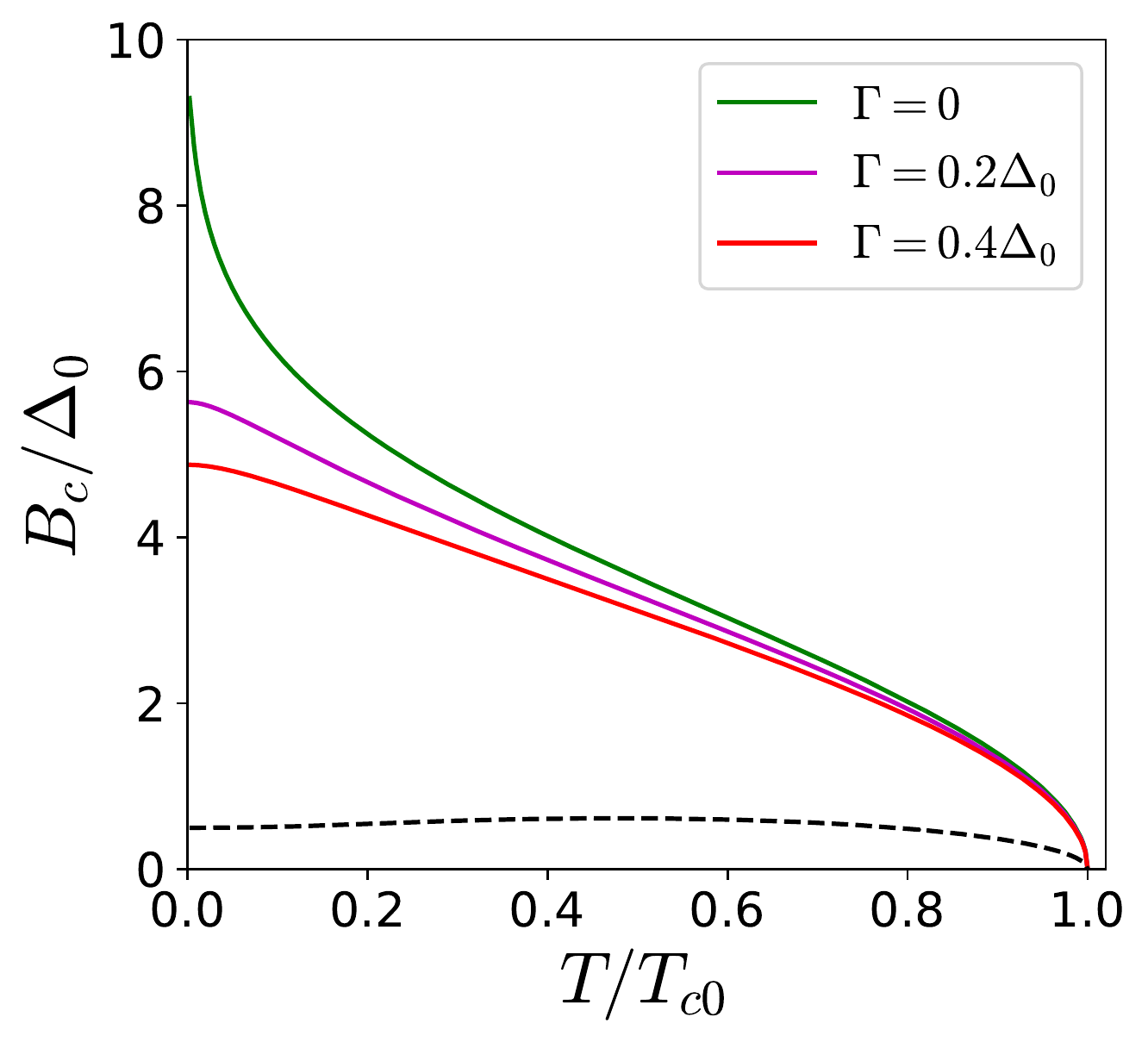}
\caption{Critical in-plane magnetic field $B_c$ versus temperature $T$ at different
intervalley scattering strengths $\Gamma$ with $\beta_{\rm so}=7\Delta_0$. The 
out-of-plane critical magnetic field is shown as the dashed line. }
\label{fig_transition_Bx}
\end{figure}

With $\Delta \rightarrow 0$ in Eq.~\eqref{g0}, one has $u =-(\varepsilon^2-\beta_{\rm
so}^2-B_x^2)$ so that Eq.~\eqref{g0} can be written as 
\begin{equation}
  g_0 = a \varepsilon \frac{\varepsilon^2-\beta_{\rm so}^2-B_x^2}{\Delta(1-\Gamma^2
  a^2B_x^2)-i\Gamma a(\varepsilon^2-B_x^2-\Delta^2)}.
\end{equation}
Since $g_0\approx 1$ under $\Delta \rightarrow 0$, one has
\begin{equation}
  a \approx
  \frac{\Delta}{(\varepsilon+i\Gamma)(\varepsilon^2-B_x^2)-\varepsilon\beta_{\rm so}^2} ,
\end{equation}
so that
\begin{equation}
  f_0(\varepsilon) =\Delta \frac{-\varepsilon(\varepsilon+i\Gamma)+\beta_{\rm
  so}^2}{(\varepsilon+i\Gamma)(\varepsilon^2-B_x^2-\beta_{\rm so}^2)+i\Gamma \beta_{\rm
  so}^2} .
\end{equation}
Using the relation \cite{Moeckli20} %Kita2015, 
\begin{equation}
  -\frac{1}{\lambda} = \ln \Big( \frac{T}{T_{c0}} \Big) +2\pi T\sum_{n=0}^{\infty} 
  \frac{1}{\omega_n} ,
\end{equation}
where $T_{c0}$ is the zero-field critical temperature, the in-plane critical magnetic
field $B_c$ can be found through the pair-breaking equation
\begin{equation}
  \ln \Big( \frac{T}{T_{c0}} \Big) = 2\pi T\sum_{n=0}^{\infty} \Big[
  \frac{\omega_n\tilde{\omega}_n +\beta_{\rm so}^2}{\tilde{\omega}_n(\omega_n^2 +B_c^2
  +\beta_{\rm so}^2) -\Gamma\beta_{\rm so}^2} -\frac{1}{\omega_n} \Big] ,
\end{equation}
with $\tilde{\omega}_n=\omega_n +\Gamma$. This expression agrees with Refs.~\cite{Ilic17,
Moeckli20}. The in-plane critical magnetic fields at different intervalley scattering
strengths are shown in Fig.~\ref{fig_transition_Bx}. 

%\begin{figure}
%\centering
%\includegraphics[width=\columnwidth]{fig_gxgz.pdf}
%\caption{(a) $g_{+,x}$ and (b) $g_{-,z}$ at different intervalley scattering rates
%  $\Gamma$ (from bottom to top: $\Gamma=0$, $0.2\Delta_0$, $0.4\Delta_0$, $\Delta_0$,
%  $2\Delta_0$) in the $K$ valley with $\beta_{\rm so}=7\Delta_0$, $B_x=2\Delta_0$ and
%  $T=0.1T_{c0}$. All curves (except the curve for $\Gamma=0$) have been offset vertically
%  for better visibility. The presented results are for the $K$ valley; for the $K'$
%  valley, the sign of $g_{-,z}$ reverses, while the sign of $g_{+,x}$ remains unchanged.}
%\label{fig_gxgz}
%\end{figure}

\end{document}